\documentclass[twocolumn]{aastex701}
\usepackage{isotope}

\graphicspath{{figs/}}
\usepackage{amsmath}
\usepackage{graphicx}
\usepackage{multirow}
\usepackage[normalem]{ulem}
\usepackage{xspace}
\usepackage{siunitx}
\newcommand\msun{M\ensuremath{_{\odot}}\xspace}  
\newcommand{\Msun}{\msun}
\newcommand{\code}[1]{\texttt{#1}}
\newcommand{\mesa}{\code{MESA}\xspace}
\newcommand{\MESA}{\mesa}

\begin{document}

\title{Impact of Nuclear Reaction Rates on Calcium Production in Population III Stars: A Global Analysis}

\author[orcid=0009-0008-8274-2521]{Qing Wang}
\affiliation{Shenzhen Key Laboratory of Research and Manufacture of High Puriny Germanium Materials and Detectors, Institute for Advanced Study in Nuclear Energy $\&$ Safety, College of Physics and Optoelectronic Engineering, Shenzhen University, Shenzhen 518060, Guangdong, People’s Republic of China}
\email{wangqing2023@email.szu.edu.cn}  

\author[orcid=0000-0002-3117-1318]{Ertao Li}
\affiliation{Shenzhen Key Laboratory of Research and Manufacture of High Puriny Germanium Materials and Detectors, Institute for Advanced Study in Nuclear Energy $\&$ Safety, College of Physics and Optoelectronic Engineering, Shenzhen University, Shenzhen 518060, Guangdong, People’s Republic of China}
\email{let@szu.edu.cn}
\correspondingauthor{let@szu.edu.cn}

\author[orcid=0009-0004-5258-9491]{Yinwen Guan}
\affiliation{Shenzhen Key Laboratory of Research and Manufacture of High Puriny Germanium Materials and Detectors, Institute for Advanced Study in Nuclear Energy $\&$ Safety, College of Physics and Optoelectronic Engineering, Shenzhen University, Shenzhen 518060, Guangdong, People’s Republic of China}
\email{2500232013@mails.szu.edu.cn}  

\author[orcid=0000-0001-5206-4661]{Zhihong Li}
\affiliation{China Institute of Atomic Energy, Beijing 102413, People’s Republic of China}
\email{zhli@ciae.ac.cn}

\author[orcid=0000-0002-3616-2680]{Jianjun He}
\affiliation{Institute of Modern Physics, Fudan University, Shanghai 200433, People’s Republic of China}
\email{hejianjun@fudan.edu.cn}

\author{Liyong Zhang}
\affiliation{School of Physics and Astronomy, Beijing Normal University, Beijing 100875, People’s Republic of China}
\email{}

\author[orcid=0000-0002-4911-0847]{Bing Guo}
\affiliation{China Institute of Atomic Energy, Beijing 102413, People’s Republic of China}
\email{guobing@ciae.ac.cn}

\author[orcid=0000-0001-7149-814X]{Youbao Wang}
\affiliation{China Institute of Atomic Energy, Beijing 102413, People’s Republic of China}
\email{ybwang@ciae.ac.cn} 

\author[orcid=0000-0001-8702-431X]{Yunju Li}
\affiliation{China Institute of Atomic Energy, Beijing 102413, People’s Republic of China}
\email{li_yunju@163.com}

\author[orcid=0000-0002-2718-9451]{Jun Su}
\affiliation{School of Physics and Astronomy, Beijing Normal University, Beijing 100875, People’s Republic of China}
\email{sujun@bnu.edu.cn}

\author[orcid=0000-0002-0029-4999]{Xiaodong Tang}
\affiliation{Institute of Modern Physics, Chinese Academy of Sciences, Lanzhou 730000, People’s Republic of China}
\email{}

\author[orcid=0000-0001-6724-2256]{Shipeng Hu}
\affiliation{Shenzhen Key Laboratory of Research and Manufacture of High Puriny Germanium Materials and Detectors, Institute for Advanced Study in Nuclear Energy $\&$ Safety, College of Physics and Optoelectronic Engineering, Shenzhen University, Shenzhen 518060, Guangdong, People’s Republic of China}
\email{husp@szu.edu.cn} 

\author[orcid=0009-0008-9198-5971]{Yu Liu}
\affiliation{Shenzhen Key Laboratory of Research and Manufacture of High Puriny Germanium Materials and Detectors, Institute for Advanced Study in Nuclear Energy $\&$ Safety, College of Physics and Optoelectronic Engineering, Shenzhen University, Shenzhen 518060, Guangdong, People’s Republic of China}
\email{2500232005@mails.szu.edu.cn} 

\author[orcid=0009-0003-3896-0269]{Dong Xiang}
\affiliation{Shenzhen Key Laboratory of Research and Manufacture of High Puriny Germanium Materials and Detectors, Institute for Advanced Study in Nuclear Energy $\&$ Safety, College of Physics and Optoelectronic Engineering, Shenzhen University, Shenzhen 518060, Guangdong, People’s Republic of China}
\email{2500231015@mails.szu.edu.cn}

\author[orcid=0009-0009-2232-4317]{Lei Yang}
\affiliation{Shenzhen Key Laboratory of Research and Manufacture of High Puriny Germanium Materials and Detectors, Institute for Advanced Study in Nuclear Energy $\&$ Safety, College of Physics and Optoelectronic Engineering, Shenzhen University, Shenzhen 518060, Guangdong, People’s Republic of China}
\email{2310456020@email.szu.edu.cn}

\author[orcid=0000-0003-3233-8260]{Weiping Liu}
\affiliation{Department of Physics, Southern University of Science and Technology, Shenzhen 518055, Guangdong, People’s Republic of China}
\email{wpliu@ciae.ac.cn}
 
\begin{abstract}
We investigate the sensitivity of calcium production to nuclear reaction rates of a $40\,\mathrm{M_\odot}$ Population III star using 1D multi-zone stellar models.
A comprehensive nuclear reaction network was constructed, and all $(p,\gamma)$ and $(p,\alpha)$ reaction rates were individually varied by a factor of 10 up and down, identifying 13 preliminary key reactions for calcium production. 
To propagate the reaction rate uncertainties on calcium production, two sets of Monte Carlo simulations were performed for these key reactions: one adopting STARLIB reaction rates and the other incorporating updated rates from recent experimental data and evaluations.
Our results show that Monte Carlo simulations using the updated rates show good agreement with the observed calcium abundance of the extremely iron-poor star SMSS J031300.36–670839.3 within the 68\% confidence interval predicted by the models. In contrast, the observed calcium abundance lies marginally outside the 68\% C.I. when using the STARLIB rates. Spearman rank-order correlation analysis and SHAP values show that the $(p,\gamma)$ and $(p,\alpha)$ reactions of ${}^{18}\mathrm{F}$ and ${}^{19}\mathrm{F}$ exhibit strong coupled effects on calcium production. These reaction-rate uncertainties need to be reduced to constrain the stellar model predictions. Our study provides insights for future nuclear physics experiments aimed at reducing reaction rate uncertainties in the nucleosynthesis of Population III Stars. 

Additionally, comparisons between $20\,\mathrm{M_\odot}$ and $40\,\mathrm{M_\odot}$ Population III stellar models confirm that the latter, with updated reaction rates, is more capable of reproducing the observed Ca abundance and [Ca/Mg] ratio. 
\end{abstract}

\section{Introduction} 
The first generation of stars, often called population III (Pop III) stars or primordial stars, are the first sites of stellar nucleosynthesis and feedback through ultraviolet (UV) radiation, supernova blast waves, and chemical enrichment in the early universe. In particular, their characteristic masses and the nature of supernova explosions are critically important in cosmic reionization and subsequent star formation~\citep{Bromm2011, Klessen2023}. Pop III stars could be very massive as a result of cooling of the metal-free primordial gas through hydrogen molecules, or they could form as lower-mass stars due to radiation feedback from growing protostars and/or disk fragmentation~\citep{Bromm2011, Ishigaki2014}.

Metal-poor stars are traditionally used as a probe of the nucleosynthetic results of Pop III supernovae. The most metal-poor stars are likely to be second-generation stars, whose chemical compositions directly record the nucleosynthetic yield of their progenitor Pop III stars~\citep{Takahashi2014}. SMSS J031300.36–670839.3 (hereafter SMSS0313–6708) is the star with the most iron-deficient discovered by the SkyMapper Southern Sky Survey so far, with [Fe/H] $< -7.1$~\citep{Keller2014} and $\le -6.53$ re-analyzed by \citet{Nordlander2017}. HE0107-5240~\citep{Christlieb2002,Christlieb2004} and HE1327-2326~\citep{Frebel2006,Frebel2008} are the other two stars with the most iron deficiency known with [Fe/H] $= -5.3$ and $-5.96$, respectively. The abundance patterns of these stars are characterized not only by extremely low iron abundance but also by enhanced intermediate-mass elements, such as carbon, nitrogen, oxygen, sodium, magnesium, and calcium.

Many efforts have been made to reproduce the abundance pattern of SMSS0313–6708 from different stellar models and nucleosynthesis processes; however, no consensus has yet been reached.~\citet{Keller2014} concluded that the star was seeded with a single low-energy supernova with an initial mass of about 60\,\Msun, and calcium production is the result of breakout from the CNO cycle during the stable hydrogen-burning phase. In addition, the observed abundance pattern does not support supernova progenitors with masses $\le$ 10\,\Msun or $\ge$ 70\,\Msun in~\citet{Keller2014}. The former eject large amounts of iron, while the latter fail to reproduce the observed carbon enhancement and simultaneously overproduce nitrogen.~\citet{Ishigaki2014} proposed an alternative scenario to explain the abundance pattern of SMSS0313–6708. They investigated supernova yields of Pop III stars and reported that the high [C/Ca] and [C/Mg] ratios and upper limits of other elemental abundances are well reproduced with the yields of core-collapse supernovae from Pop III 25\,\Msun or 40\,\Msun stars. The best-fit models assume that the explosions undergo extensive mixing and fallback, leaving behind a black hole remnant. In these models, calcium is produced by static/explosive O burning and incomplete Si burning in the supernova/hypernova, rather than originating from the hot-CNO cycle during the main-sequence phase, as suggested by~\citet{Keller2014,Bessell2015}.~\citet{Takahashi2014} demonstrated that in the non-rotating 80\,\Msun model, calcium is produced at the base of the hydrogen-burning shell owing to the breakout of the hot-CNO cycle after the carbon-burning stage. The predictions of the model are compatible with the observed abundance pattern, while the origin of calcium differs slightly from that proposed by~\citet{Keller2014}. However, the calcium production is much lower in 30\,\Msun -- 70\,\Msun non-rotating models, and models with masses $\ge$ 100\,\Msun are rejected due to the overabundance of magnesium.~\citet{Clarkson2018} proposed the $i$-process triggered by H-ingestion events (see also~\citealt{Takahashi2014} and ~\citealt{Clarkson2020}) as a potential nucleosynthesis mechanism for calcium production in Pop III stars. Their results show that magnesium originates from He burning and calcium from the $i$-process. Although the model reproduces the overall abundance pattern of SMSS0313–6708, it overproduces sodium by more than 2.5 dex, which may rule out this scenario.

Using single-zone nucleosynthesis calculations with the temperature and density from their 80Mled stellar model,~\citet{Clarkson2020} identified the reaction $^{19}$F(p, $\gamma$)$^{20}$Ne as the most important breakout path, allowing catalytic material to leak out of the CNO cycle towards the NeNa cycle under hydrogen-burning conditions in Pop III stars.~\citet{Zhang2022} also refer to this reaction as a breakout reaction from the ``warm'' CNO cycle. It would cause irreversible flow from the CNO to the NeNa range because back-processing through $^{22}\mathrm{Ne}(p, \alpha)^{19}\mathrm{F}$ (similar to the case of $^{18}\mathrm{O}(p, \alpha)^{15}\mathrm{F}$) is not energetically possible~\citep{Wiescher1999}. Hot-CNO breakout during shell Hydrogen-burning has little impact on calcium abundance due to the limited time before collapse~\citep{Clarkson2020}.~\citet{Clarkson2020} also examined their four stellar models, namely 15Mled, 80Mled, 140Mled and 60Mled-l, without experiencing H–He interactions. Their analysis reveals that 98\% of the calcium in the hydrogen-burning shell originates from the main sequence in their Pop III stellar models. The maximum amount of calcium produced by hydrostatic hydrogen burning in these models (except for 140Mled) is approximately $0.8$--$2$ dex lower than the calcium abundance observed in SMSS0313–6708. However, they noted that increasing the reaction-rate ratio of $^{19}\mathrm{F}(p,\gamma)/^{19}\mathrm{F}(p,\alpha)$ by roughly an order of magnitude relative to the NACRE compilation~\citep{Angulo1999} may allow the models to reproduce the observed calcium abundance.

\citet{deBoer21} revised $^{19}\mathrm{F}(p, \gamma)^{20}\mathrm{Ne}$ and its competing channel $^{19}\mathrm{F}(p, \alpha)^{16}\mathrm{O}$ via the phenomenological R-matrix approach. They concluded that revising the breakout reaction reduces the efficiency of calcium production under hydrogen-burning conditions in Pop III stars. ~\citet{Williams21} measured the strength of the 323 keV resonance in the $^{19}\mathrm{F}(p, \gamma)^{20}\mathrm{Ne}$ reaction using inverse kinematics techniques, resulting in the $^{19}\mathrm{F}(p,\gamma)/^{19}\mathrm{F}(p,\alpha)$ ratio being increased by a factor of two in $T < 0.1~\mathrm{GK}$ relative to the compilation by~\citet{Angulo1999}. \citet{Zhang2022} performed a direct experimental measurement of the reaction $^{19}\mathrm{F}(p, \gamma)^{20}\mathrm{Ne}$ to a very low energy of 186 keV on the Jinping Underground laboratory for Nuclear Astrophysics (JUNA) accelerator at the China JinPing Underground Laboratory (CJPL) and reported a key resonance at 225 keV. As a result, the revised reaction rate is enhanced by a factor of 5.4-7.4 at around $0.1~\mathrm{GK}$ relative to the recommended rate in \citet{Angulo1999}. This enhancement suggests more efficient calcium production due to leakage from the “warm” CNO cycle in 40\,\Msun Pop III stellar models, compatible with the calcium abundance observed in SMSS0313–6708. More recently,~\citet{Su25} indirectly measured the cross section of $^{19}\mathrm{F}(p,\alpha\gamma)^{16}\mathrm{O}$ using the Trojan Horse Method (THM), suggesting a reduced strength of the 11 keV resonance. The revised $^{19}\mathrm{F}(p,\gamma)^{20}\mathrm{Ne}$ reaction rate was reported to be enhanced by a factor of 4.6–6.4 at $0.1~\mathrm{GK}$ relative to the recommended rate in~\citet{Angulo1999} (see references therein), resulting in a lower $^{19}\mathrm{F}(p,\gamma)/^{19}\mathrm{F}(p,\alpha)$ ratio below $0.3~\mathrm{GK}$ compared to that reported by~\citet{Zhang2022}. They drew a conflicting conclusion that increasing the $^{19}\mathrm{F}(p,\gamma)/^{19}\mathrm{F}(p,\alpha)$ ratio relative to the NACRE compilation~\citep{Angulo1999} is insufficient to explain the observed calcium abundance in stars such as SMSS0313–6708, even using the $^{19}\mathrm{F}(p,\gamma)^{20}\mathrm{Ne}$ rate of~\citet{Zhang2022}, based on their 20\,\Msun and 25\,\Msun Pop III stellar models.

The possibility of a break-out from the ``warm” CNO cycles depends not only on the abundance of $^{19}\mathrm{F}$, but also on the thermonuclear rates of the reaction $^{19}\mathrm{F}(p,\gamma)^{20}\mathrm{Ne}$ and the competing reaction $^{19}\mathrm{F}(p,\alpha)^{16}\mathrm{O}$ ~\citep{Wiescher1999,Zhang2022}. Therefore, reactions before and beyond  $^{19}\mathrm{F}$ may also determine the calcium abundance. Previous studies mainly focused on the $^{19}\mathrm{F}(p,\gamma)^{20}\mathrm{Ne}$ channel and its competing $^{19}\mathrm{F}(p,\alpha)^{16}\mathrm{O}$ channel. A question that deserves careful reconsideration is whether the calcium abundance observed in SMSS0313–6708 can be reproduced within a global analysis framework using reaction rates available prior to~\citet{Zhang2022} in 40\,\Msun Pop III stellar models.

Motivated by the above question, we performed extensive large-scale Monte Carlo simulations like~\citet{Fields2016,Fields2018} to study the impact of thermonuclear reaction rates on calcium production in Pop III stars. In this work, we first systematically identified the nuclear reactions that most strongly influence calcium production by varying all $(p,\gamma)$ and $(p,\alpha)$ reaction rates individually in the nuclear network by a factor of 10 up and down in 40,\Msun Pop III stellar models. Subsequently, we carried out two grids of Monte Carlo stellar simulations, one including the preliminary key reactions with STARLIB rates~\citep{Iliadis2010a,Sallaska2013} and the other including the preliminary key reactions with updated rates from the latest evaluations and experimental data, to propagate the reaction-rate uncertainties. In order to respond to previous concerns, we also compare 20\,\Msun with 40\,\Msun Pop III stellar models using different $^{19}\mathrm{F}(p,\gamma)^{20}\mathrm{Ne}$ rates in the end.

This paper is novel in two main aspects. First, we performed large-scale Monte Carlo simulations to propagate reaction-rate uncertainties directly within Population III stellar models, different from post-processing schemes. This method accounts for coupled reaction effects on calcium production, allowing for a direct comparison with the calcium abundance observed in the most iron-deficient star currently known, SMSS0313–6708. Second, we apply
XGBoost (eXtreme Gradient Boosting)~\citep{Chen2016} and SHAP (SHapley Additive exPlanations)~\citep{Lundberg2017,Lundberg2020} analysis to interpret the results of Monte Carlo simulations, providing a systematic way to disentangle the complex dependencies between nuclear reaction rates and calcium production.

In Section~\ref{sec:Stellar Model}, we describe the input physics of our stellar models and discuss the characteristics of the baseline model. In Section~\ref{sec:Sensitivity Test}, we present the results of sensitivity tests. In Section~\ref{sec:Monte Carlo Simulations}, we describe the Monte Carlo method and introduce the reaction-rate sets used in Monte Carlo simulations. In Section~\ref{sec:Results}, we present the results of the Monte Carlo simulations. In Section~\ref{sec:Impacts of the updated rates}, we compare the results of 20\,\Msun with 40\,\Msun Pop III stellar models. In Section~\ref{sec:Discussion and Conclusion}, we summarize our results and present our conclusions.

\section{\label{sec:Stellar Model}Stellar Model} 
We use the \MESA stellar evolution code~\citep[r24.03.01,][]{Paxton2010,Paxton2013,Paxton2015,Paxton2018,Paxton2019,Jermyn2023} to evolve 40\,\Msun models, beginning with the primordial composition taken from~\citet{Pitrou2020}. The stellar models are based on the template \code{20M\_pre\_ms\_to\_core\_collapse} in the \code{test\_suite} folder, with modified input parameters. We do not include stellar rotation in our models. Stellar mass loss is also neglected, since Pop III stars are expected to have no efficient winds due to pulsations~\citep{Baraffe2001, Krtika2006}.

\subsection{\label{subsec:Input Physics}Input Physics} 
We focus on Ca production during hydrogen burning, including both the core and shell hydrogen-burning phases. Following the nuclear reaction networks adopted in~\citet{Brinkman2021,Takahashi2014,Clarkson2020}, as well as \code{mesa\_80.net}, we construct a nuclear reaction network consisting of 137 species up to Cu (Table~\ref{tab:Nuclear reaction network}), comprising more than 1300 strong and weak reactions. The network includes possible proton-capture reactions associated with hot-CNO breakout, but does not explicitly model neutron-capture processes, such as the $i$-process~\citep{Clarkson2018}.

\begin{table}[!htbp]
	\centering
	\caption{\label{tab:Nuclear reaction network}The isotopes in the nuclear reaction network used in MESA simulations.}
	\begin{tabular*}{0.4\textwidth}{@{\extracolsep{\fill}}c c c c}
		\hline\hline
		Element & $A$ & Element & $A$ \\
		\hline
		n   & 1        & P   & 28--33 \\
		H   & 1--2     & S   & 31--36 \\
		He  & 3--4     & Cl  & 32--37 \\
		Li  & 6--7     & Ar  & 35--40 \\
		Be  & 7, 9, 10 & K   & 36--41 \\
		B   & 8--13    & Ca  & 39--44 \\
		C   & 10--14   & Sc  & 41--45 \\
		N   & 11--15   & Ti  & 44--50 \\
		O   & 13--18   & V   & 47--51 \\
		F   & 16--19   & Cr  & 48--54 \\
		Ne  & 18--22   & Mn  & 51--55 \\
		Na  & 20--23   & Fe  & 52--58 \\
		Mg  & 22--26   & Co  & 55--56 \\
		Al  & 25--27   & Ni  & 56--59 \\
		Si  & 26--30   & Cu  & 59     \\
		\hline
	\end{tabular*}
\end{table}

We use the Ledoux criterion to establish the location of the convective boundaries with the mixing length parameter of $\alpha_{\mathrm{MLT}} = 1.5$~\citep{Brinkman2019, Brinkman2021}, and the time-dependent convection (TDC) scheme for convection. Convective boundary mixing (CBM) is included in the form of semiconvection, overshooting, and thermohaline mixing. For semiconvection, we use the prescription of~\citet{Langer1985} with an efficiency parameter $\alpha_{\mathrm{sc}} = 0.5$ following~\citet{Clarkson2020}. For overshoot mixing, we use the step-overshoot scheme with $f = 0.35$ and $f_{\mathrm{0}} = 0.01$ at the core hydrogen-burning stage. For thermohaline mixing, we use $\alpha_{\mathrm{th}} = 1.0$.

For spatial resolution controls, we use the \code{MESA}’ \code{mesh\_delta\_coeff}, $\delta_{\mathrm{mesh}}$, with a default value of 1.0, which accounts for gradients in structural quantities to determine whether a cell should be split or merged. We also apply a minimum number of total cells using \code{max\_dq}$ = 10^{-3}$, which is the maximum fraction of the total mass contained within each cell. Specifically, we set \code{mesh\_delta\_coeff\_for\_highT = 0.5} during the Si-burning phase until the onset of iron-core collapse, i.e., the pre-supernova (Pre-SN) stage. The above controls result in an average of approximately 2,000–2,200 cells from the zero-age main sequence (ZAMS) to the terminal-age main sequence (TAMS), around 2,800 cells at core Ne-depletion, and approximately 6,000–7,000 cells during the Si-burning phase until the Pre-SN. We use the same \code{varcontrol\_target}, $w_t$, of $3\times10^{-4}$ from the pre-main sequence (PMS) until the Pre-SN. For changes in composition, we further limit the time steps using \code{dX\_nuc\_drop\_limit} with a value of $1 \times 10^{-3}$, which specifies the largest allowed decrease in mass fraction due to nuclear burning or mixing for species with $X_i \ge 10^{-4}$.

The JINA Reaclib nuclear reaction rate library v2.2~\citep{Cyburt2010} is used as the basis for rate variations. Inverse rates are calculated directly from the forward rates using detailed balance. Our stellar models are mainly divided into three parts as follows. First, the baseline stellar model evolves from the PMS until the Pre-SN. Evolving stellar models until core Ne-depletion (core $X_{\mathrm{c}}(^{20}\mathrm{Ne}) \lesssim 10^{-3}$) provides a good approximation and meets the goals of this work, as described in Section~\ref{subsec:Baseline Stellar Model}. Second, in Section~\ref{sec:Sensitivity Test}, all reactions $(p, \gamma)$ and $(p, \alpha)$ in the nuclear reaction network are individually varied by a factor of 10 up and down to identify the preliminary key reactions responsible for Ca production. Each stellar model evolves from the PMS to the TAMS, defined when the core $X_{\mathrm{c}}(^{1}\mathrm{H}) \lesssim 10^{-2}$. The stellar models with modified preliminary key reaction rates will continue to evolve until core Ne-depletion. Third, in Section~\ref{sec:Monte Carlo Simulations}, we employ the Monte Carlo method to calculate the uncertainty in Ca production, taking into account all the key reactions determined. These Monte Carlo models evolve until core Ne-depletion. 
\subsection{\label{subsec:Evolution of the Baseline}Evolution of the Baseline $\mathrm{40\,M_{\odot}}$ Pop III Stellar Model}
In this section, we present the characteristics of the baseline 40\,\Msun stellar model, which evolves from the PMS until the Pre-SN with the JINA Reaclib v2.2, as described in Section~\ref{subsec:Input Physics}.
\begin{figure}[!htbp]
    \centering
    \includegraphics[width=\columnwidth]{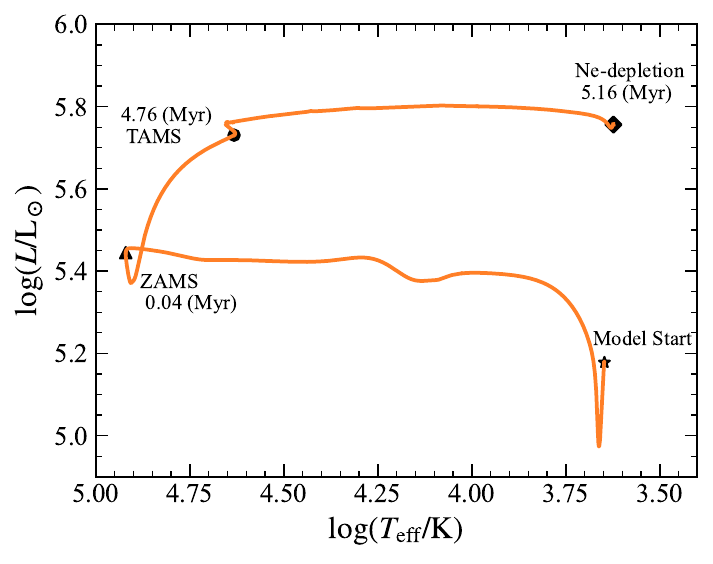}
    \caption{\label{fig:HR}(Color online) HRD of the baseline 40\,\Msun stellar model from the PMS until the Pre-SN. The star symbol denotes the beginning of the stellar model, the triangle indicates the ZAMS, the circle denotes the TAMS, and the diamond represents core Ne-depletion. Ages at these stages are annotated. The subsequent evolution after core Ne-depletion until the Pre-SN shows only minor changes in luminosity and is therefore not depicted.}
\end{figure} 

\begin{figure}[!htbp]
	\centering
	\includegraphics[width=\columnwidth]{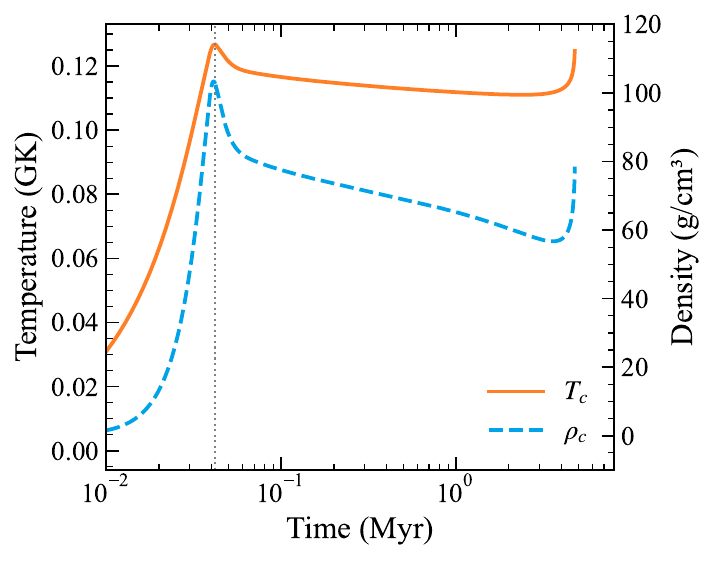}
	\caption{\label{fig:T_Rho}(Color online) The evolution of the central temperature (orange) and density (blue) up to the TAMS for the baseline model. The gray dotted line represents the ZAMS, at which temperature and density reach the peaks.}
\end{figure}
Figure~\ref{fig:HR} shows the Hertzsprung–Russell diagram (HRD) of the baseline stellar model. The beginning of the stellar model, the ZAMS, the TAMS, and core Ne-depletion are indicated by symbols. Minor changes in luminosity are not shown between core Ne-depletion and the Pre-SN. At the ZAMS, the model has a luminosity and an effective temperature of $\log(L/\mathrm{L_\odot}) \simeq 5.44$ and $\log(T_{\mathrm{eff}}/\mathrm{K}) \simeq 4.92$, respectively. The model spends $\simeq 0.04~\mathrm{Myr}$ on the PMS and $\simeq 4.72~\mathrm{Myr}$ on the MS. The burning phases between the TAMS and core Ne-depletion last for $\simeq 0.4~\mathrm{Myr}$, while those from core Ne-depletion until the Pre-SN last only for $\simeq 0.275~\mathrm{yr}$.

\begin{figure}[!htbp]
	\centering
	\includegraphics[width=\columnwidth]{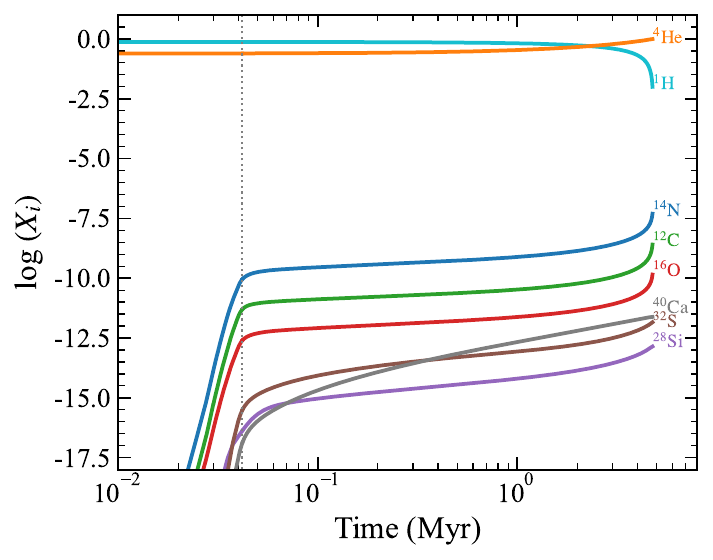}
	\caption{\label{fig:Abundance_Tams}(Color online) Central mass fraction of main isotopes as a function of time up to the TAMS for the baseline model. The gray dotted line represents the ZAMS. }
\end{figure}

\begin{figure}[!htbp]
	\centering
	\includegraphics[width=\columnwidth]{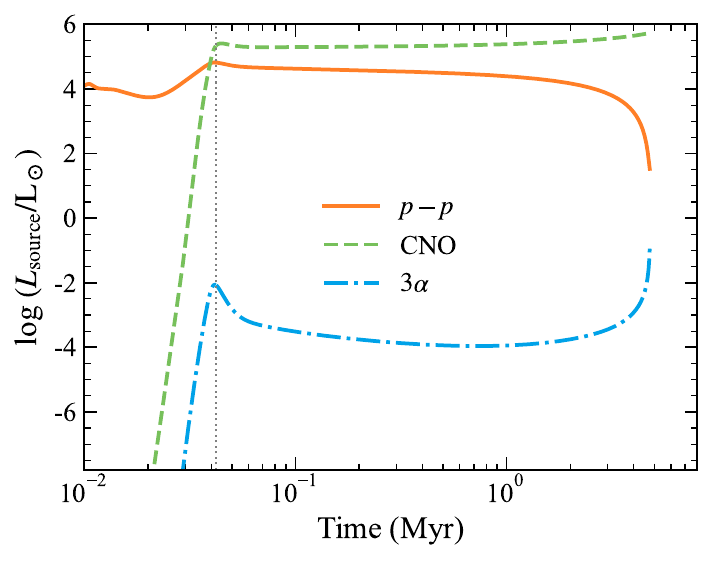}
    	\caption{\label{fig:HR_source}(Color online) The luminosities of the $p-p$ chain reactions (orange solid line), the CNO reactions (green dashed line), and the $3\alpha$ process (blue dot-dashed line) as functions of time up to the TAMS for the baseline model. The gray dotted line represents the ZAMS, around which CNO reactions overtake $p-p$ chain reactions as the dominant luminosity source.}
\end{figure}
Massive Pop III stars begin H burning via the $p–p$ chain reactions with the primordial Big Bang composition in the absence of CNO seeds. However, these reactions are not efficient enough to provide the energy needed to power the star and maintain thermal equilibrium. The star contracts in order to keep the equilibrium. As the central temperature and density rise, the $3\alpha$ process is eventually initiated, creating a small amount of ${}^{12}\mathrm{C}$ in the core that serves as a catalyst for the activation of the CNO cycle~\citep{Clarkson2020,Zhang2022}. As shown in Figure~\ref{fig:T_Rho}, the central temperature and density increase during this contraction phase and reach their peak values at the ZAMS. The corresponding evolution of the main central isotopes is shown in Figure~\ref{fig:Abundance_Tams}. Following the onset of the $3\alpha$ process, the central ${}^{12}\mathrm{C}$ mass fraction increases, which allows the CNO cycle to operate efficiently. As the CNO reactions become more efficient, their contribution to nuclear energy generation increases rapidly, and the luminosity quickly overtakes that of the $p$–$p$ chain reactions, making them the dominant energy source (Figure~\ref{fig:HR_source}).  At around 0.1 GK, catalytic material leaks out of the CNO cycle towards the NeNa cycle through ${}^{19}\mathrm{F}(p,\gamma)^{20}\mathrm{Ne}$, after which the reaction flow proceeds almost unhindered toward the stable nucleus ${}^{32}\mathrm{S}$, the double magic nucleus ${}^{40}\mathrm{Ca}$, and other heavier isotopes~\citep{Zhang2022}. Other breakout reactions, e.g, those $\alpha$-induced reactions on ${}^{15}\mathrm{O}$ and ${}^{18}\mathrm{Ne}$ nuclei, are open up yet at this temperature~\citep{Wiescher1999}.  

\begin{figure}[!htbp]
	\centering
	\includegraphics[width=\columnwidth]{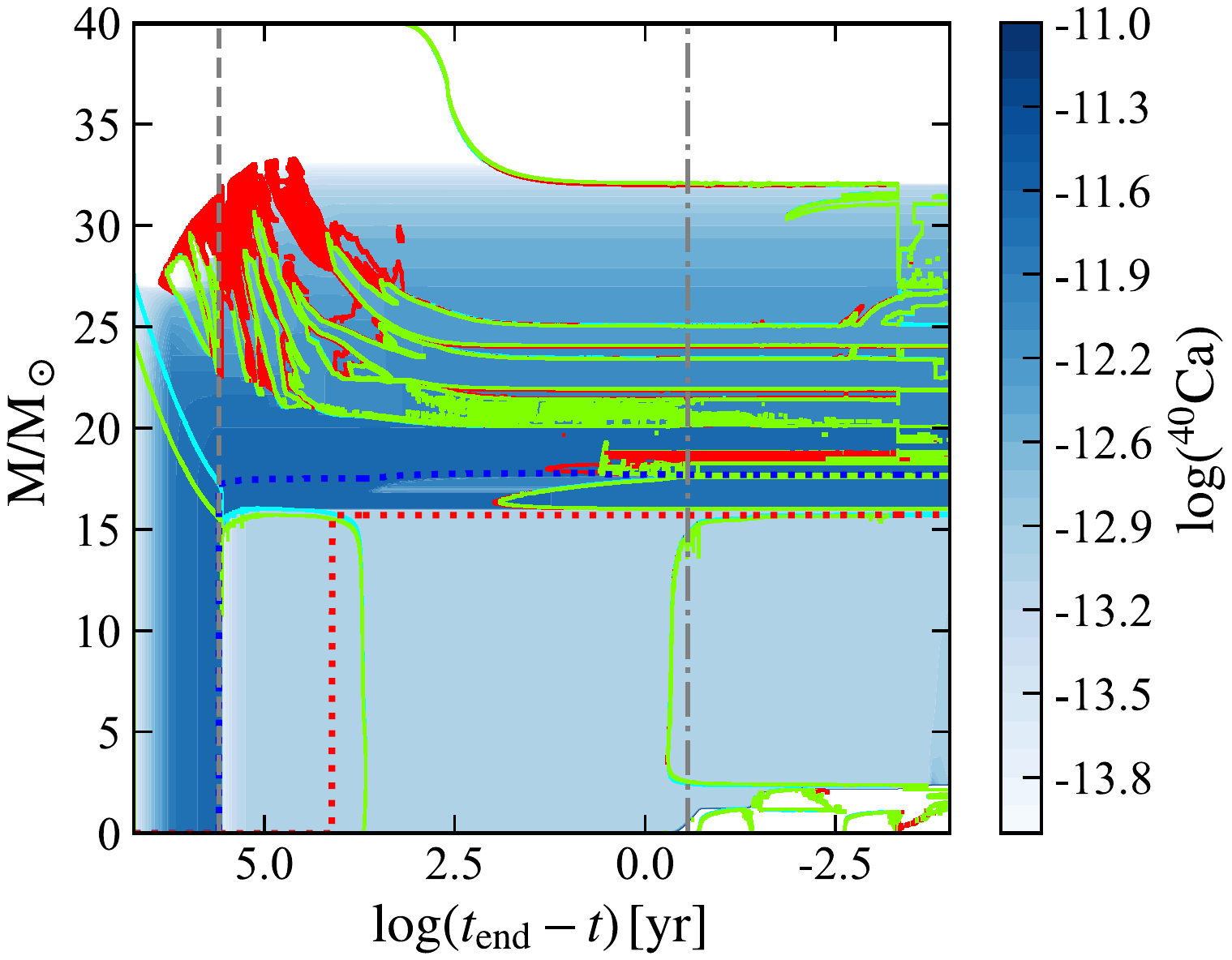}
	\caption{\label{fig:a09_DC_1.5_Kippenhahn}(Color online) Kippenhahn diagram for the baseline 40\,\Msun stellar model. The horizontal axis shows the time left until the Pre-SN, while the vertical axis denotes the mass coordinate. The green, cyan, and red-hatched areas correspond to convective, overshoot, and semi-convective regions, respectively. The blue dotted line indicates the H-depleted core, or He core, where the H mass fraction is below 0.01, and the He mass fraction is above 0.01. The red dotted line indicates the He-depleted core, or CO core, where the He mass fraction is below 0.01. The blue color scale shows the ${}^{40}\mathrm{Ca}$ mass fraction. The vertical gray dashed line represents the TAMS, while the dot-dashed line represents core Ne-depletion.}
\end{figure}

Figure~\ref{fig:a09_DC_1.5_Kippenhahn} shows the Kippenhahn diagram for the baseline 40\,\Msun stellar model, illustrating the internal structural evolution and the distribution of ${}^{40}\mathrm{Ca}$ mass fraction from the PMS until the Pre-SN. The horizontal axis denotes the remaining time until the Pre-SN, while the vertical axis shows the mass coordinate. During the MS, the H-burning core is convective. This convective mixing continuously transports fresh H fuel from the entire convective region into the center, where H burning is strongly concentrated, thereby making the core almost chemically homogeneous~\citep{Lamers2017}. At the TAMS as indicated by the vertical gray dashed line, for ${}^{40}\mathrm{Ca}$ abundance, the central mass fraction is $\log ({}^{40}\mathrm{Ca}) = -11.62$, while the average mass fraction of the whole star is $\log ({}^{40}\mathrm{Ca}) = -11.86$, with Ca being dominated by ${}^{40}\mathrm{Ca}$. As the star evolves, the convective core shrinks due to the lower opacity of He compared to H. After the exhaustion of H in the core, H burning proceeds in a shell, while the He core below undergoes advanced nuclear burning phases that further modify the abundances~\citep{Heger2010}. Specifically during He burning, ${}^{40}\mathrm{Ca}$ is destroyed by neutron-capture reactions~\citep{Lugaro2018,Brinkman2021}, and the presence of the convective He-burning core thus leads to reduced ${}^{40}\mathrm{Ca}$ abundance in the core, as shown in Figure~\ref{fig:a09_DC_1.5_Kippenhahn}.

\subsection{\label{subsec:Baseline Stellar Model}Supernova Explosions and Comparison with Observations}
Fallback in core-collapse supernovae is considered a major ingredient for explaining abundance anomalies in metal-poor stars~\citep{Chan2018}, known as SMSS0313-6708. In this scenario, most of the material below the He core of a supernova progenitor is accreted onto the black hole, and the ejected material experiences a variety of thermodynamic conditions and mixing processes, finally seeding SMSS0313–6708~\citep{Zhang2022}. In this work, we do not simulate the supernova explosion, but follow the two assumptions of~\citet{Takahashi2014}: The abundance distribution in the outer layers of the progenitor is not significantly changed by explosive nucleosynthesis, and the weak explosion only expels the matter in the outer region of the star. 

The mass fraction of Ca, as well as the abundance ratios [Ca/H] and [Ca/Mg], are calculated and compared with the observed values $\log (\mathrm{Ca}) = -11.4$~\citep{Zhang2022}, [Ca/H] $= -7.2$, and [Ca/Mg] $= -2.9$~\citep{Keller2014}, respectively. 

\begin{equation} 
	\label{eq2}
	X_{\mathrm{hhe}} = \frac{\int_{M_{\mathrm{b}}}^{M_{\mathrm{surface}}} {X} dm}{\int_{M_{\mathrm{b}}}^{M_{\mathrm{surface}}}{dm}}, 
\end{equation}
where $X_{\mathrm{hhe}}$ is the average mass fraction of a given element, $X$ is the mass fraction of a given element in the mass element $dm$, $M_{\mathrm{surface}}$ is the mass coordinate corresponding to the surface of the star, and $M_{\mathrm{b}}$ is the mass coordinate corresponding to the base of the He shell. In order to mimic the ejection fraction $f$~\citep{Takahashi2014}, We calculate abundance values for different mass coordinates of the base of the He shell where ${}^{4}\mathrm{He}$ mass fraction, $X_{\mathrm{b}}(^{4}\mathrm{He})$, satisfies $0.01$--$0.1$. 

\begin{figure*}[!htbp]
\centering
\includegraphics[width=0.9\columnwidth]{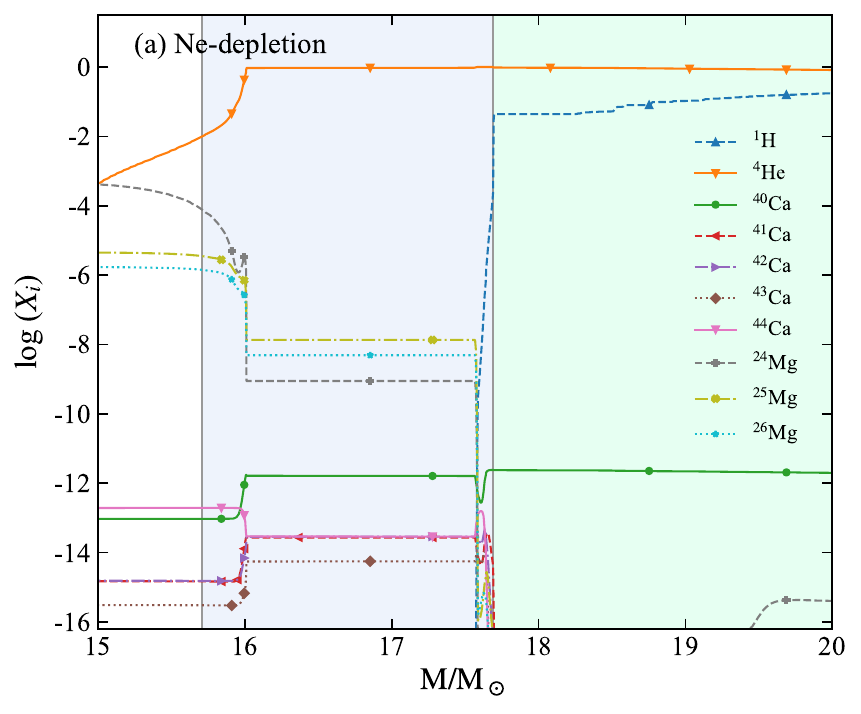}
\includegraphics[width=0.9\columnwidth]{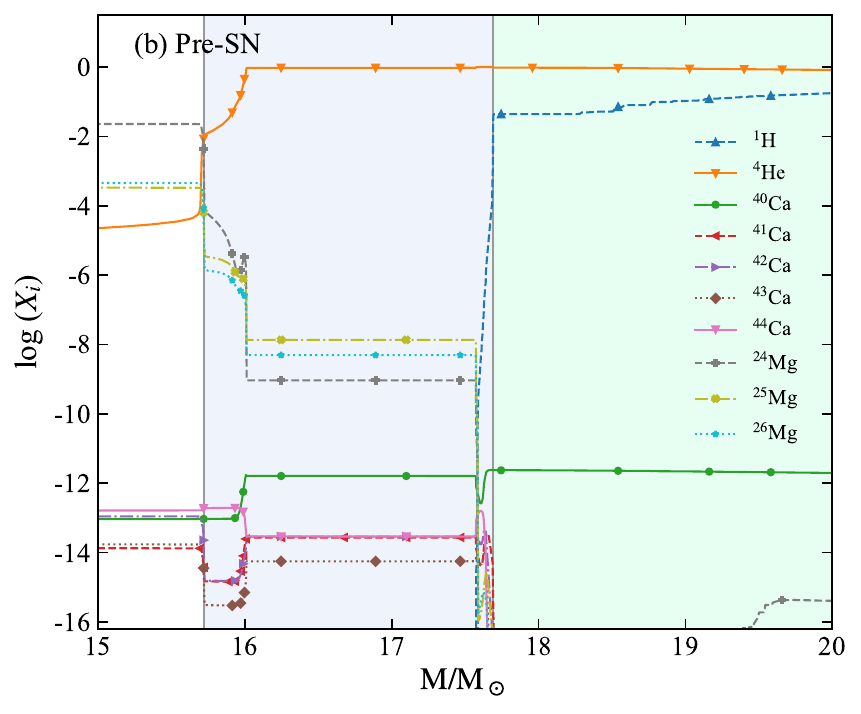}
\caption{\label{fig:Abundance_Profile}(Color online) The profiles of the stable Ca and Mg isotopic abundances at core Ne-depletion (a) and the Pre-SN (b) for the baseline model. The blue and green areas represent the He shell and H envelope, respectively.}
\end{figure*}

\begin{figure}[!htbp]
	\centering
	\includegraphics[width=0.9\columnwidth]{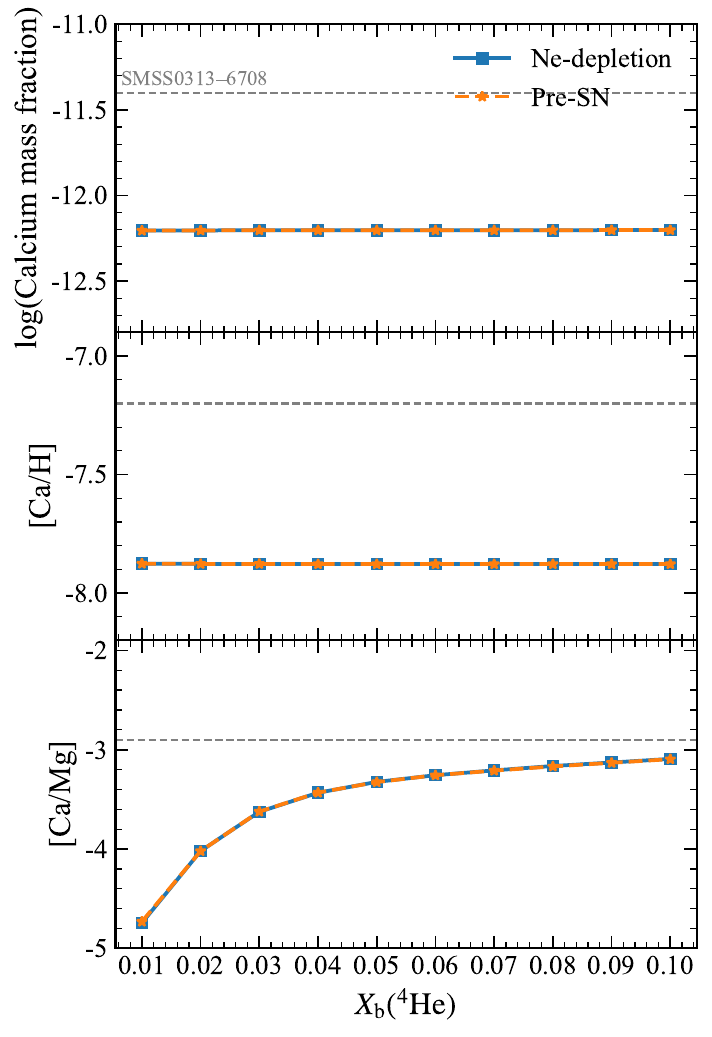}
	\caption{\label{fig:baseline}(Color online) The Ca mass fraction (top), [Ca/H] (middle), and [Ca/Mg] (bottom) as a function of $X_{\mathrm{b}}(^{4}\mathrm{He})$ at core Ne-depletion and the Pre-SN for the baseline model. One can see that the values are almost the same between core Ne-depletion and the Pre-SN. The gray dashed lines represent the observations. The [Ca/Mg] ratio is sensitive to the choice of the mass coordinate defining the base of the He shell, in contrast to the constant Ca mass fraction and [Ca/H].}
\end{figure}

The solar standardized values of
\begin{eqnarray}
[i/j] = \mathrm{log} \Bigl( \frac{ X_{i} }{ X_{j} } \Bigl) - \mathrm{log} \Bigl( \frac{ X_{i} }{ X_{j}} \Bigl)_{\odot} \label{eq1}
\end{eqnarray}
are calculated using solar values by~\citet{Asplund2009}. We do not include any dilution from the interstellar medium (ISM)~\citep{Clarkson2020}; therefore, our estimates represent the maximum Ca mass fraction, [Ca/H], and [Ca/Mg] values obtained from simulations in this work.   

Figure~\ref{fig:Abundance_Profile} shows the profiles of the stable Ca ($T_{1/2}({}^{41}\mathrm{Ca}) \simeq 99.4~\mathrm{kyr}$~\citep{Brinkman2021}) and Mg isotopic abundances at core Ne-depletion and the Pre-SN for the baseline model. We found that the isotopic abundances above the CO core remain largely unchanged between these two stages. The abundances of Mg isotopes are low in the H envelope and show a decreasing trend outside the CO core. Figure~\ref{fig:baseline} compares the baseline results with the observations. We found that the values are almost the same between core Ne-depletion and the Pre-SN. The Ca mass fraction and [Ca/H] remain nearly constant, while [Ca/Mg] increases with $X_{\mathrm{b}}(^{4}\mathrm{He})$. The [Ca/Mg] ratio is sensitive to the choice of the mass coordinate defining the base of the He shell. We concluded that evolving the stellar models up to core Ne-depletion provides a good approximation for comparing the observations. 

\section{\label{sec:Sensitivity Test}Sensitivity Test} 

\begin{figure*}[!htbp]
\centering
\includegraphics[width=\textwidth]{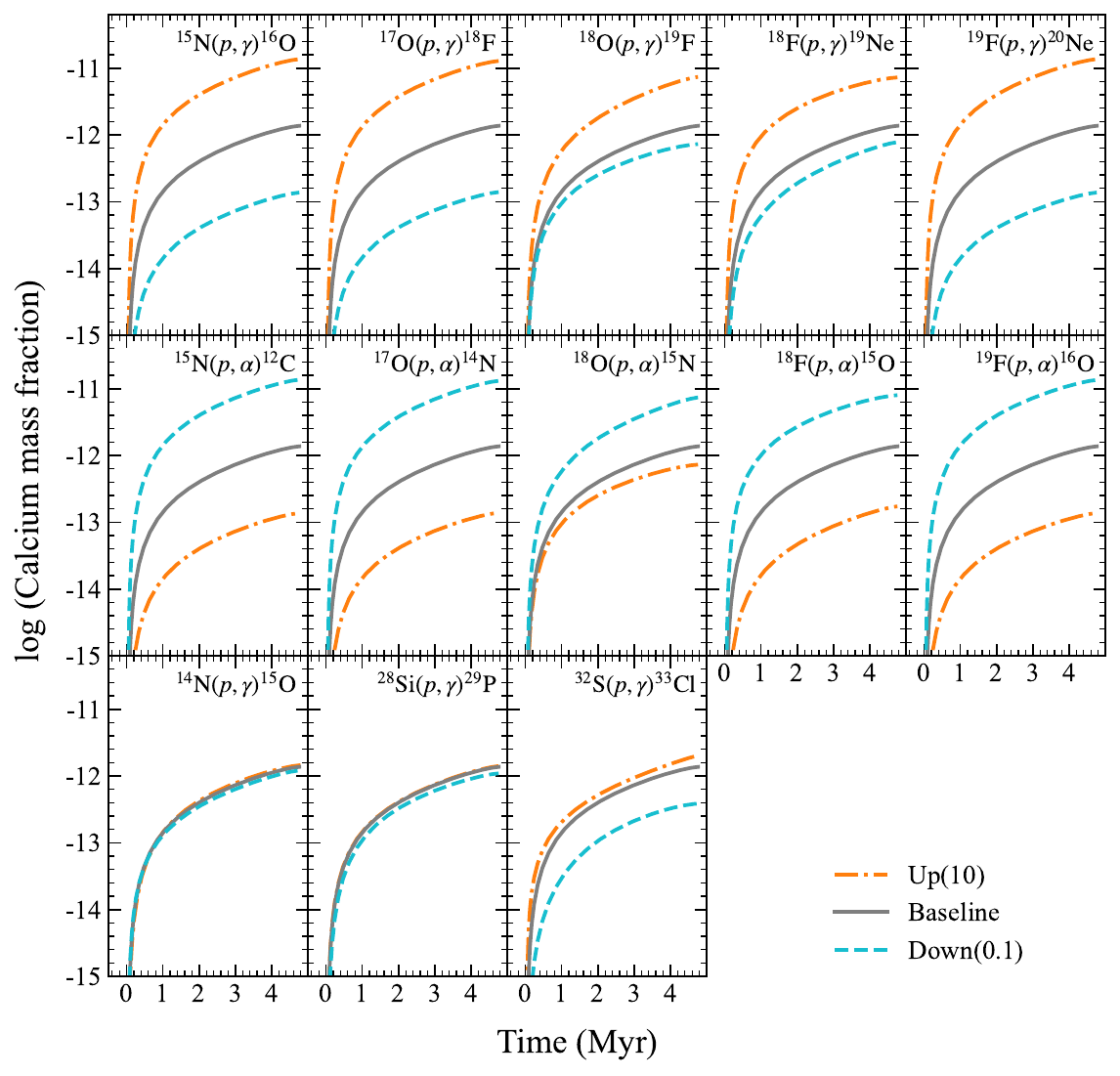}
\caption{\label{fig:Sensitivity_Test}(Color online) Evolution of the average Ca mass fraction of the whole star induced by a factor of 10 up (orange dot-dashed lines) and down ( blue dashed lines) for the 13 preliminary key reactions. Models evolve from the PMS to the TAMS. Gray solid lines represent the baseline result.} 
\end{figure*}

In this section, we present the results of the sensitivity tests. As discussed above, Ca is primarily produced during the MS, and this work focuses on the (p, $\gamma$) and (p, $\alpha$) reactions that affect Ca production. Evolving stellar models up to the TAMS is appropriate for investigating sensitivity to reaction rates, but it is not able to directly compare the observations. We have varied all (p, $\gamma$) and (p, $\alpha$) reaction rates individually in the nuclear network (see also Table~\ref{tab:Nuclear reaction network}) by a factor of 10 up and down, based on JINA Reaclib v2.2. As a result, 13 reactions that primarily affect Ca production were identified.

Figure~\ref{fig:Sensitivity_Test} illustrates the evolution of the average Ca mass fraction of the whole star, $X_{\mathrm{s}}(\mathrm{Ca})$, from the PMS to the TAMS, resulting from varying the 13 preliminary key reaction rates by a factor of 10 up and down. The $(p,\gamma)$ and $(p,\alpha)$ reactions of $^{15}$N, $^{17}$O, and $^{19}$F have the largest impact on $X_{\mathrm{s}}(\mathrm{Ca})$ ($\sim 2$~dex at the TAMS, defined as the difference between the upper and lower limits), followed by $^{18}$F$(p,\alpha)^{15}$O ($\sim 1.7$~dex), 
$^{18}$F$(p,\gamma)^{19}$Ne, $^{18}$O$(p,\gamma)^{19}$F, and $^{18}$O$(p,\alpha)^{15}$N ($\sim 1.0$~dex), 
$^{32}$S$(p,\gamma)^{33}$Cl ($\sim 0.7$~dex), $^{28}$Si$(p,\gamma)^{29}$P ($\sim 0.1$~dex), and $^{14}$N$(p,\gamma)^{15}$O ($\sim 0.08$~dex).

\begin{figure}[!htbp]
	\centering
	\includegraphics[width=\columnwidth]{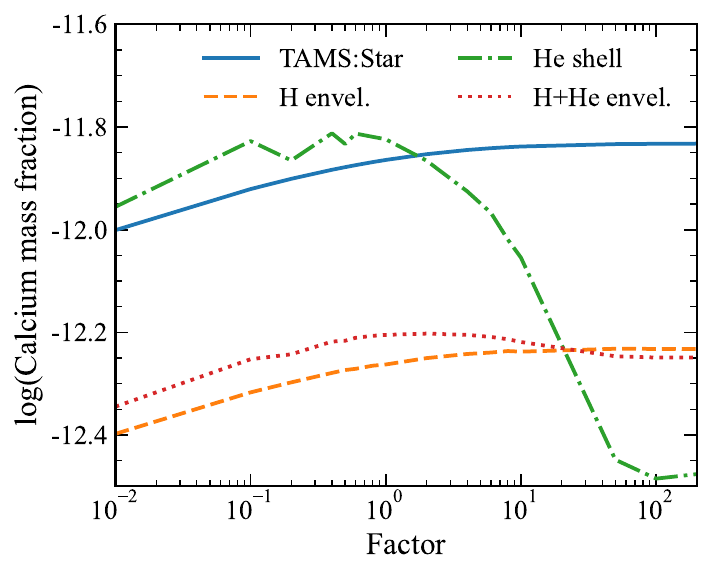}
	\caption{\label{fig:n14_pg_o15_factor}(Color online) The Ca mass fraction as a function of variation factor for $^{14}$N$(p, \gamma)^{15}$O rate. The blue solid line represents the average of the Ca mass fraction for the whole star at the TAMS. The dot-dashed, dashed, and dot-dotted lines are the average of the Ca mass fraction for He shell, H envelope, and their combination at core Ne-depletion.}
\end{figure}

Afterwards, the stellar models with modified rates of the preliminary key reactions were evolved until core Ne-depletion. At this stage, the variations in $X_{\mathrm{hhe}}(\mathrm{Ca})$ are similar to those of $X_{\mathrm{s}}(\mathrm{Ca})$ at the TAMS, except for the $^{14}$N$(p,\gamma)^{15}$O case, where the variation is reduced from $\sim 0.08$~dex to $\sim 0.03$~dex (see the Table~\ref{tab:ca_sensitivity} in the Appendix~\ref{sec:Calcium Table}). The average Ca mass fraction in the He shell, $X_{\mathrm{he}}(\mathrm{Ca})$, is lower than the corresponding baseline value even when the $^{14}$N$(p,\gamma)^{15}$O reaction rate is increased by a factor of 10. To further investigate the effect of the reaction rate, another 16 models were calculated by varying the $^{14}\mathrm{N}(p,\gamma)^{15}\mathrm{O}$ reaction rate by factors ranging from 0.01 to 250. Figure~\ref{fig:n14_pg_o15_factor} demonstrates that $X_{\mathrm{he}}(\mathrm{Ca})$ is non-monotonic with the reaction rate and the $^{14}\mathrm{N}(p,\gamma)^{15}\mathrm{O}$ reaction has a significant impact on the Ca abundance distribution in the He shell.

\section{\label{sec:Monte Carlo Simulations}Monte Carlo Simulations}
\subsection{\label{subsec:Monte Carlo Method}Monte Carlo Method}
\begin{figure}[!htbp]
\centering
\includegraphics[width=\columnwidth]{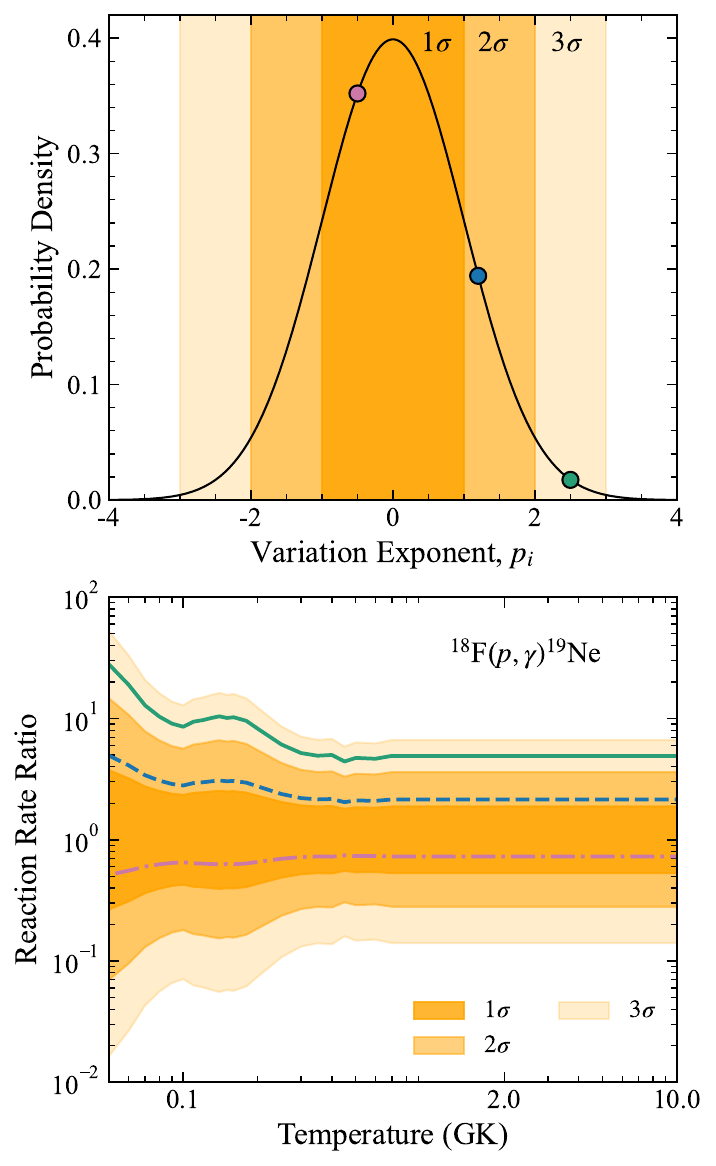}
\caption{\label{fig:tdmc}(Color online) (Top) Standard normal distribution of reaction rate variation exponent $p_i$. Three points of $p_i = -0.5, 1.2$, and 2.5 are also drawn from left to right. 
(Bottom) The resulting reaction rates compared to the median for the $^{18}$F($p$, $\gamma$)$^{19}$Ne reaction rate. The red dot-dashed, blue dashed, and green solid lines represent the samplings from $p_i = -0.5, 1.2$, and $2.5$, respectively.}
\end{figure}

The STARLIB rate library~\citep{Iliadis2010a,Sallaska2013} provides the median nuclear reaction rate, $r_{med}$, and the associated factor uncertainty, $f.u.$ (for a 68\% coverage probability), at temperature ranges from 0.001 GK to 10 GK~\citep{Sallaska2013}. All rates in STARLIB also follow a lognormal probability distribution at each temperature $T$.
\begin{equation}
	f(r)=\frac{1}{\sqrt{2\pi}\sigma r}\, 
	e^ {(-(ln(r) - \mu)^2/2 \sigma^2},~\mathrm{for~0<r<\infty},
\end{equation}
where r is the reaction rate. The lognormal function has two parameters $\mu$ and $\sigma$. Following~ \citet{Longland2010,Sallaska2013}, we can get that:
The median rate, $r_{med}$, is related to the location parameter $\mu$:
\begin{equation}
	r_{med}= e^\mu =\sqrt{r_{low}r_{high}}
\end{equation}
while the factor uncertainty, $f.u.$, of the rate is related to the spread parameter $\sigma$:
\begin{equation}
	f.u.= e^\sigma =\sqrt{r_{high}/r_{low}}
\end{equation}
So the low rate and high rate are given by:
\begin{equation}
    r_{low}=r_{med}/f.u. \quad r_{high}= r_{med}f.u.
\end{equation}
In each Monte Carlo model $i$, a random variation factor $p_{ij}$ is assigned to each reaction rate~\citep{Longland2012,Sallaska2013,Fields2016}:
\begin{equation}\label{eq:tdmc}
	r_{ij} = e^{\mu} (e^{\sigma})^{p_{ij}} = r_{\rm med} \, (f.u.)^{p_{ij}},
\end{equation}

where $p_{ij}$ is drawn from a normal distribution with mean of zero and
standard deviation of unity for the $j$-th sampled reaction rate. When the rate variation factor $p_{ij}=0$, the median rate $r_{med}$ can be recovered. Since the factor uncertainty $f.u.$ varies with temperature, a sampled rate distribution maintains the temperature dependence of the rate uncertainty. For a more detailed discussion of the sampling procedure used in Monte Carlo nucleosynthesis studies, the reader is referred to~\citet{Longland2012}.

In brief, following~\citep{Psaltis2025}, we illustrate the above sampling procedure in Figure~\ref{fig:tdmc} using the example of $^{18}$F(p, $\gamma$)$^{19}$Ne. In the top panel, the red, blue, and green points correspond to three rate variation factors, $p_i = -0.5$, $1.2$, and $2.5$, drawn from the standard normal distribution. In the bottom panel, these points are assigned to reaction rates according to Equation~\ref{eq:tdmc}, and then compared to the median, corresponding to the solid lines dotted with red dots, blue dashed and green solid lines, respectively. The reaction rate ratio is not constant but follows the temperature dependence of the rate uncertainty.

\subsection{\label{subsec:reaction rates}Reaction Rates for Monte Carlo Simulations}
The 13 reactions adopted in the Monte Carlo simulations are listed in Table~\ref{tab:sampled_rates}, together with two sets of reaction rates used for samples, denoted as SetA and SetB. All reaction rates in SetA are taken from the STARLIB database. Based on SetA, we constructed SetB by updating reaction rates from the latest evaluations and experimental data. Specifically, the rates of $^{18}$F$(p,\gamma)^{19}$Ne, $^{28}$Si$(p,\gamma)^{29}$P, and $^{32}$S$(p,\gamma)^{33}$Cl remain identical to those in SetA. The $^{18}$F$(p,\alpha)^{15}$O reaction rate reported by \citet{Kahl2021,Portillo2023} still carries large uncertainties; therefore, we also retain the STARLIB rate for this reaction. 

\begin{deluxetable}{clllcc}[!htbp]
\tablewidth{1.0\linewidth}
\tablecaption{\label{tab:sampled_rates}Reaction Rates for Monte Carlo Simulations}
\tablehead{
		\colhead{} & 
		\colhead{} & 
		\multicolumn{2}{c}{Reference} \\
		\cline{3-4} 
		\colhead{Reaction Index}  &
		\colhead{Reaction} & 
		\colhead{SetA} & 
		\colhead{SetB} 
}
\startdata
	1&	$^{14}$N$(p, \gamma)^{15}$O      & \texttt{im05} (Exp.)  & \texttt{nac2} (Exp.) \\
	2&	$^{15}$N$(p, \gamma)^{16}$O      & \texttt{nacr} (Exp.)  & \texttt{nac2} (Exp.) \\
	3&	$^{15}$N$(p, \alpha)^{12}$C      & \texttt{nacr} (Exp.)  & \texttt{nac2} (Exp.) \\
	4&	$^{17}$O$(p, \gamma)^{18}$F      & \texttt{bu15} (Exp.)  & \texttt{luna} (Exp.) \\
	5&	$^{17}$O$(p, \alpha)^{14}$N      & \texttt{bu15} (Exp.)  & \texttt{luna} (Exp.) \\
	6&	$^{18}$O$(p, \gamma)^{19}$F      & \texttt{mc13} (MC)    & \texttt{mc26} (MC)   \\
	7&	$^{18}$O$(p, \alpha)^{15}$N      & \texttt{mc13} (MC)    & \texttt{li24} (Exp.) \\
	8&	$^{18}$F$(p, \gamma)^{19}$Ne     & \texttt{mc10} (MC)    & \texttt{mc10} (MC)   \\
	9&	$^{18}$F$(p, \alpha)^{15}$O      & \texttt{mc10} (MC)    & \texttt{mc10} (MC)   \\
	10&	$^{19}$F$(p, \gamma)^{20}$Ne     & \texttt{nacr} (Exp.)  & \texttt{juna} (Exp.) \\
	11&	$^{19}$F$(p, \alpha)^{16}$O      & \texttt{nacr} (Exp.)  & \texttt{nacr} (Exp.) \\
	12& $^{28}$Si$(p, \gamma)^{29}$P     & \texttt{mc10} (MC)    & \texttt{mc10} (MC)   \\
	13& $^{32}$S$(p, \gamma)^{33}$Cl     & \texttt{mc10} (MC)    & \texttt{mc10} (MC)   \\
\enddata
\tablecomments{Experimental reaction rates with approximate uncertainties are labeled ``Exp.", whereas those with statistically rigorous, Monte Carlo–based uncertainties are labeled ``MC".\texttt{im05}---\citet{Imbriani05}, \texttt{nacr}---\citet{Angulo1999}, \texttt{bu15}---\citet{Buckner15}, \texttt{mc13}---\citet{Sallaska2013}, \texttt{mc10}---\citet{Iliadis2010a}, \texttt{nac2}---\citet{Xu2013}, \texttt{luna}---\citet{Rapagnani25}, \texttt{mc26}---\citep{Buckner2012,Sallaska2013,Best2019,Bruno2019,Pantaleo2021}, \texttt{li24}---\citet{Li2024}, \texttt{juna}---\citet{Zhang2022}. The reaction rates of SetA are all taken from STARLIB.}
\end{deluxetable}

\section{\label{sec:Results}Results}
In this section, we present the results of Monte Carlo simulations. We have evolved two grids of 40\,\Msun Monte Carlo stellar models from the PMS until Ne depletion, with each grid comprising 1,000 models. The first grid consists of stellar models evolved with reaction rates sampled from SetA, while the second grid uses reaction rates sampled from SetB; other input physics are the same as the baseline model. The total computational expense amounts to $\simeq 1.2~\mathrm{M}$ CPU hours on our servers.

\subsection{\label{subsec:SROC}Spearman Rank-order Correlation and SHAP}
We use the average Ca mass fraction above the base of the He shell, where $X_{\mathrm{b}}(^{4}\mathrm{He})~= 0.01$, to perform analysis in this section. As discussed in Section~\ref{subsec:Baseline Stellar Model}, $X_{\mathrm{hhe}}(\mathrm{Ca})$ is not sensitive to $X_{\mathrm{b}}(^{4}\mathrm{He})$.

To determine the reaction rates that have the largest impact on Ca production, we conduct Spearman rank-order correlation analysis. The Spearman correlation coefficient, $r_{\mathrm{s}}$, is defined as the Pearson correlation coefficient of the ranked variables. It can be calculated as~\citep{Bliss2020},
\begin{equation}
r_{\mathrm{s}} =
\frac{
\sum_{i=1}^{N}
\left(R(p_{i}) - \overline{R(p)}\right)
\left(R(X_{i}) - \overline{R(X)}\right)
}{
\sqrt{
\sum_{i=1}^{N}
\left(R(p_{i}) - \overline{R(p)}\right)^{2}
}
\sqrt{
\sum_{i=1}^{N}
\left(R(X_{i}) - \overline{R(X)}\right)^{2}
}
},
\label{eq:spearmanrho}
\end{equation}
where $N$ is the total number of stellar models in one set. $R(p_i)$ and $R(X_i)$ correspond to the ranks of the reaction rate variation factors and final Ca abundances, respectively. $\overline{R(p)}$ and $\overline{R(X)}$ are the average ranks. A Spearman correlation coefficient of $r_{\mathrm{s}}=+1$ corresponds to a perfectly monotonically increasing relationship, while $r_{\mathrm{s}}=0$ indicates the absence of a monotonic correlation. In contrast, $r_{\mathrm{s}}=-1$ corresponds to a perfectly monotonically decreasing relationship.

\begin{figure}[!htbp]
	\centering
	\includegraphics[width=\columnwidth]{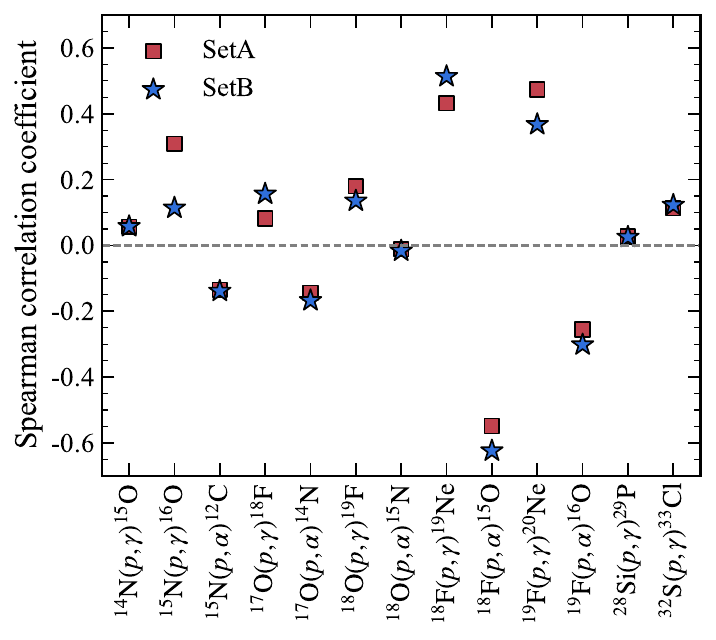}
	\caption{\label{fig:Spearman_Ca_HHe}(Color online) Spearman correlation coefficients obtained from Monte Carlo simulations of the 13 independently sampled nuclear reaction rates for SetA (red squares) and SetB (blue stars).}
\end{figure}

Figure~\ref{fig:Spearman_Ca_HHe} compares the Spearman correlation coefficients for SetA and SetB. In SetB,  $^{15}$N$(p, \gamma)^{16}$O becomes less correlated with Ca production, with $r_{\mathrm{s}}$ changing from $0.31$ to $0.11$. The absolute values of the Spearman correlation coefficients, $|r_{\mathrm{s}}|$, for $^{18}$F$(p, \gamma)^{19}$Ne ($r_{\mathrm{s}}=0.51$) and $^{18}$F$(p, \alpha)^{15}$O ($r_{\mathrm{s}}=-0.62$) increase, whereas $|r_{\mathrm{s}}|$ for $^{19}$F$(p, \gamma)^{20}$Ne ($r_{\mathrm{s}}=0.37$) decreases. In both SetA and SetB, $|r_{\mathrm{s}}|$ for the $(p,\gamma)$ and $(p,\alpha)$ reactions of $^{18}$F and $^{19}$F are larger than $0.25$.

While the Spearman rank-order correlation analysis provides a straightforward measure of the monotonic relationship between individual reaction rates and Ca production, it does not account for potential nonlinear effects or interactions among multiple reactions. To complement Spearman rank-order correlation analysis, we employ XGBoost to model the simultaneous influence of all reaction rates on Ca production. We then use SHAP to interpret the XGBoost models (see~\citet{Lundberg2019} for details). The contribution of each reaction rate to the predicted Ca production is then quantified using SHAP values. 

\begin{figure}[!htbp]
	\centering
	\includegraphics[width=\columnwidth]{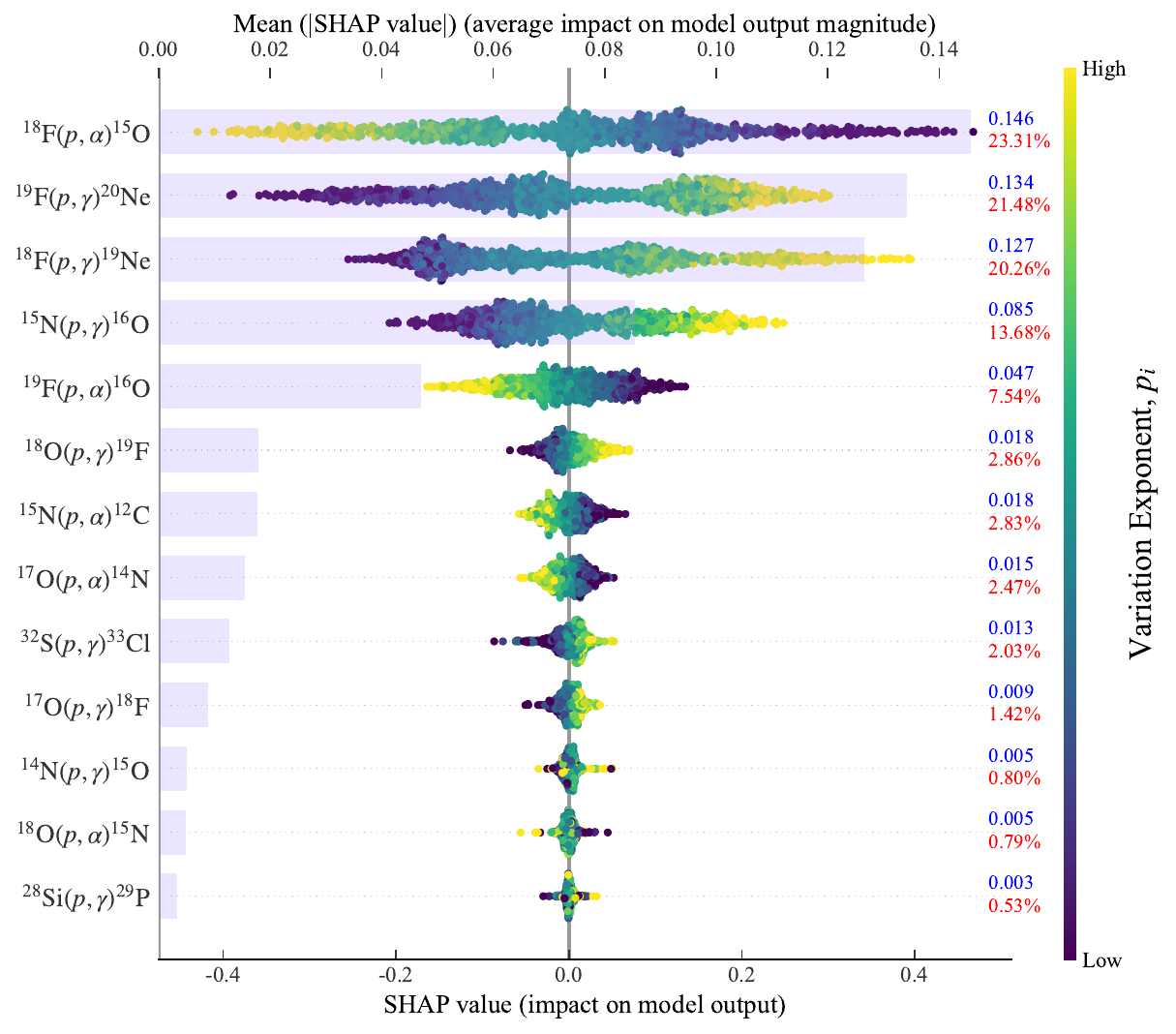}\\
    \includegraphics[width=\columnwidth]{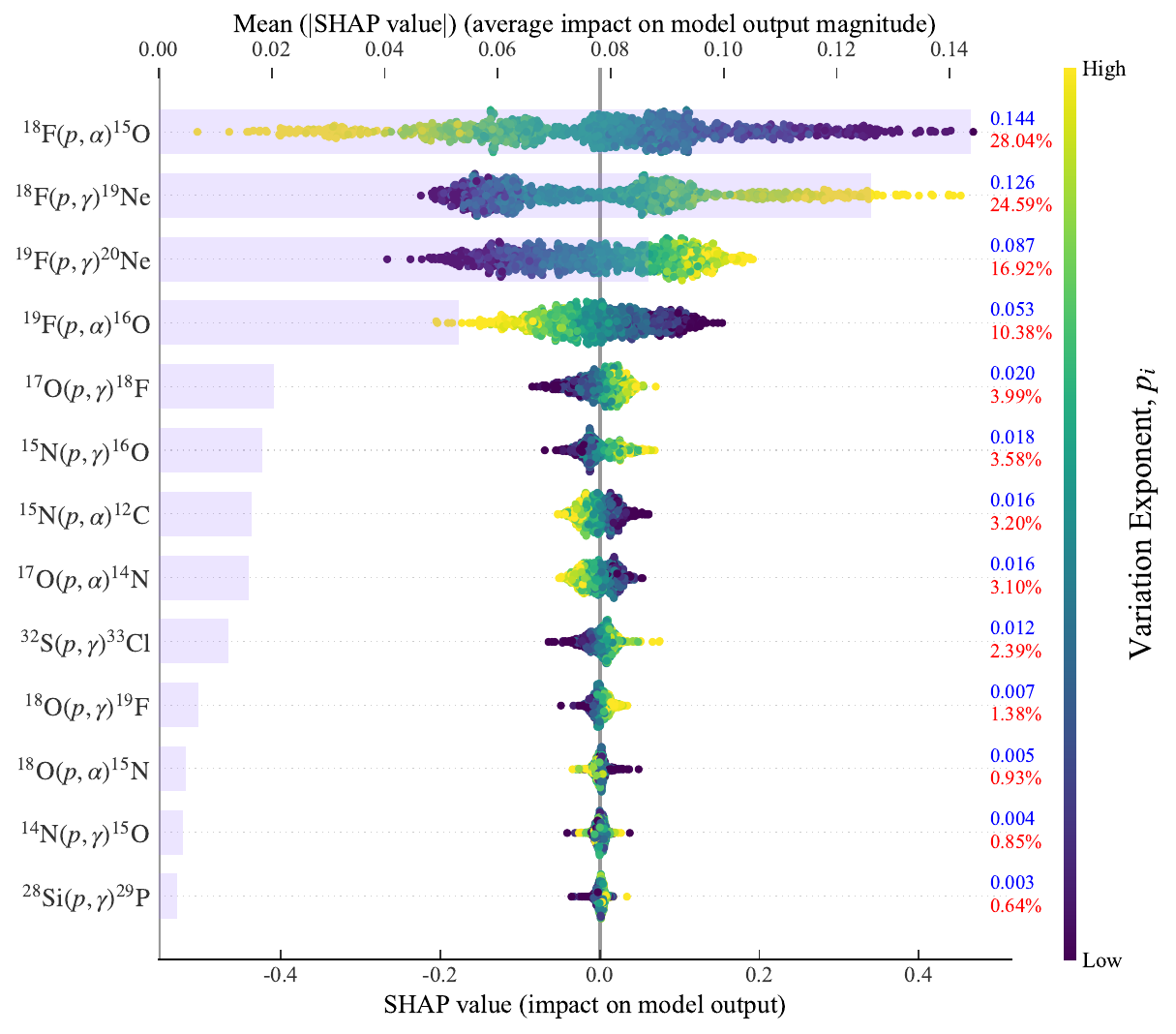}
	\caption{\label{fig:SHAP_values_and_percentages}(Color online) SHAP summary plot of the 13 reactions for SetA (Top) and SetB (Bottom). The higher the SHAP value of a reaction rate, the higher the contribution to the abundance of Ca. Each dot is created for each sample, $p_i$, of that reaction, from low (dark purple) to high (yellow). The mean of the absolute SHAP values (light blue bars) is also plotted. One can see the order of the reactions which influence the production of Ca.}
\end{figure}

\begin{figure*}[!hbp]
\centering
    \vspace{-31pt}
	\gridline{
     \hfill
		\fig{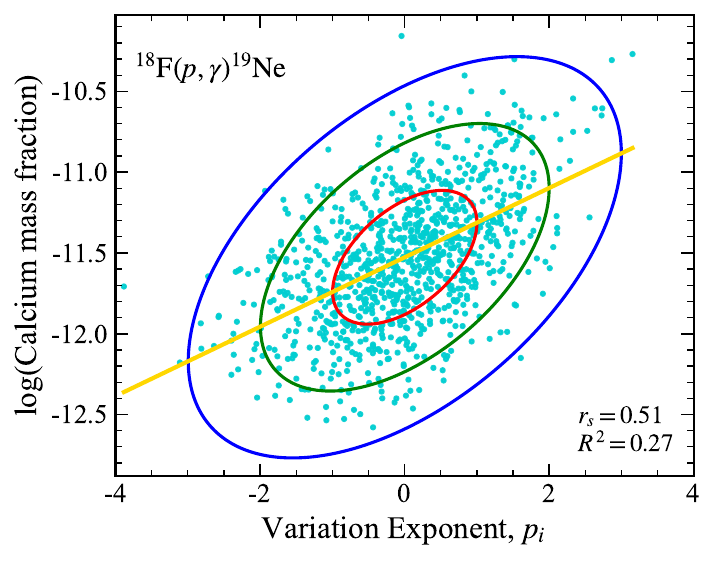}{0.38\textwidth}{}
		\fig{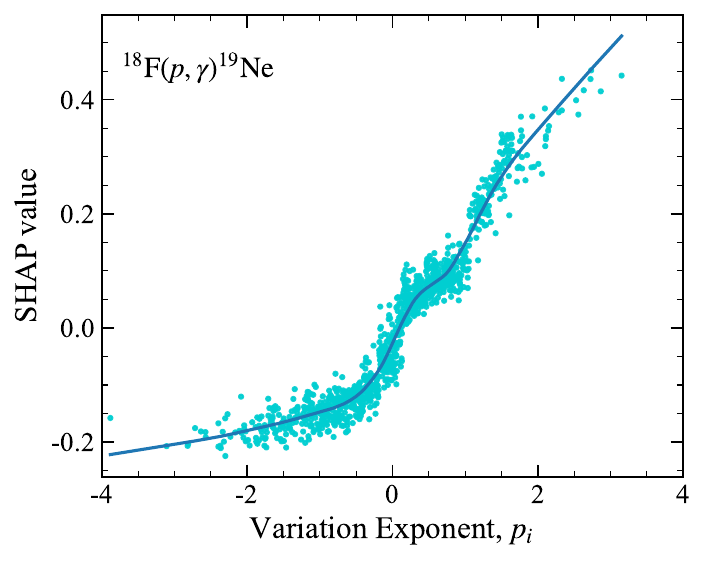}{0.38\textwidth}{}
     \hfill
	}
	\vspace{-31pt}
    \gridline{	
    \hfill
		\fig{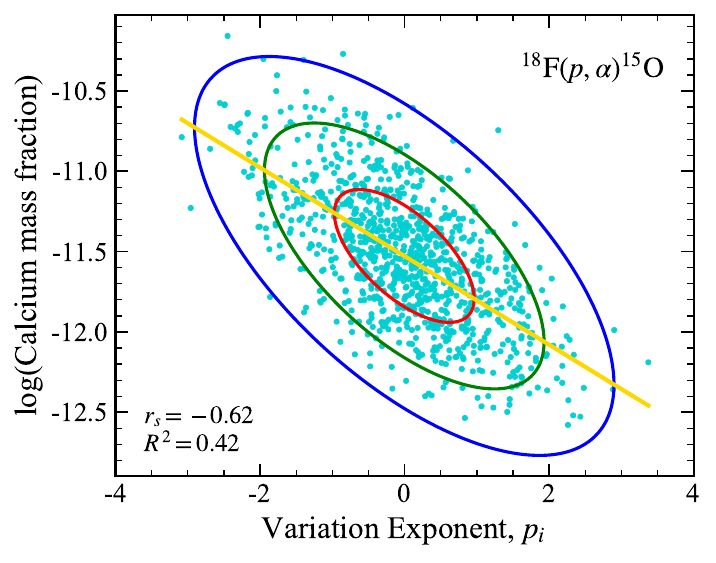}{0.38\textwidth}{}
		\fig{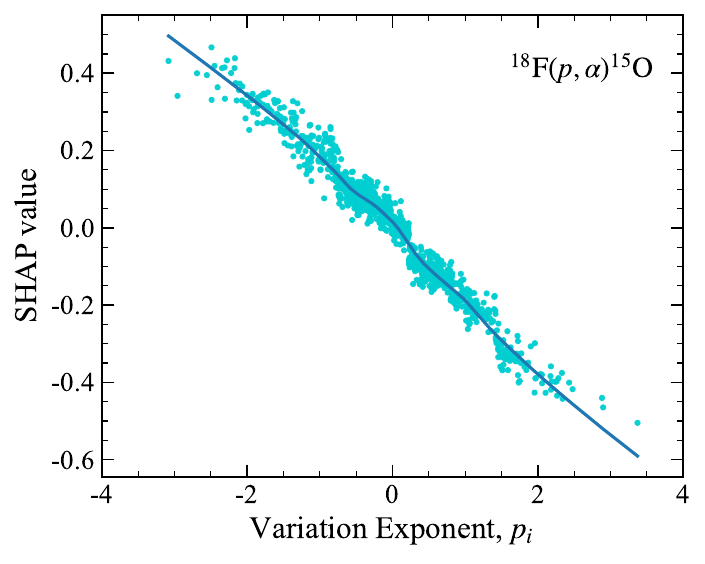}{0.38\textwidth}{}
    \hfill
	}
	\vspace{-31pt}
    \gridline{
    \hfill
		\fig{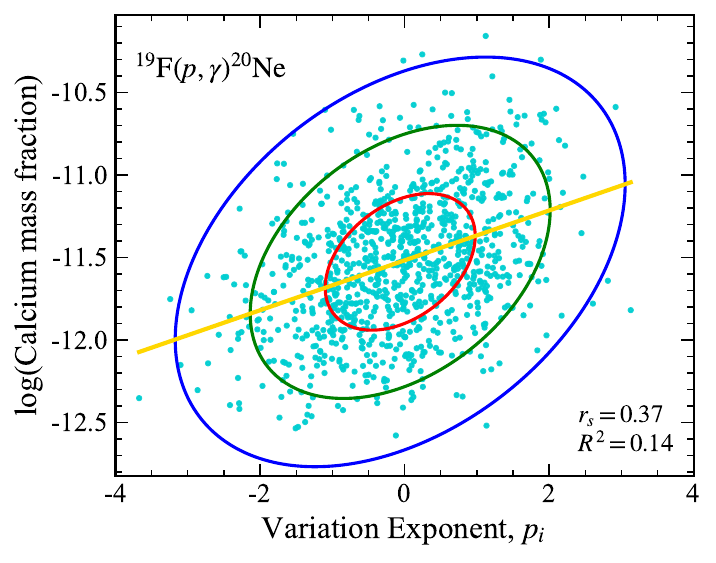}{0.38\textwidth}{}
		\fig{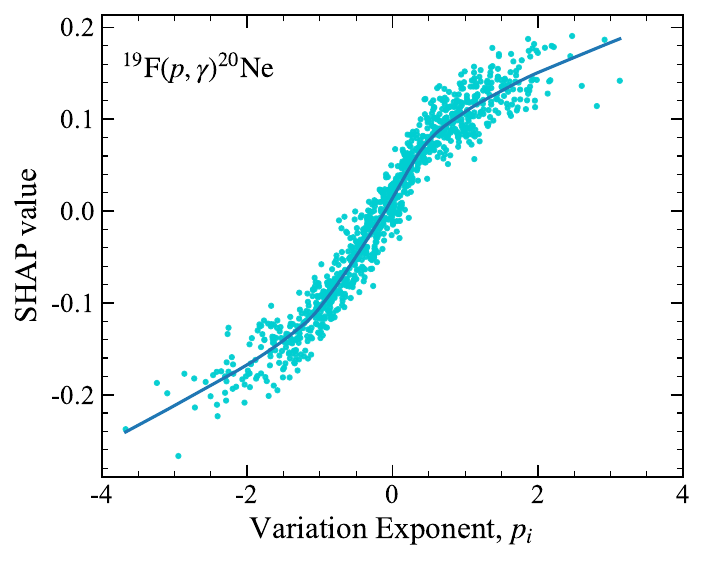}{0.38\textwidth}{}
    \hfill
	}
	\vspace{-31pt}
    \gridline{		
    \hfill
		\fig{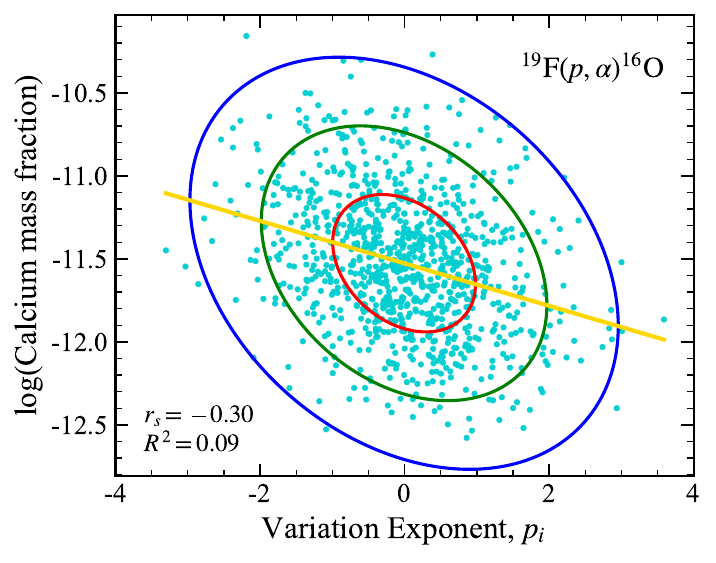}{0.38\textwidth}{}
		\fig{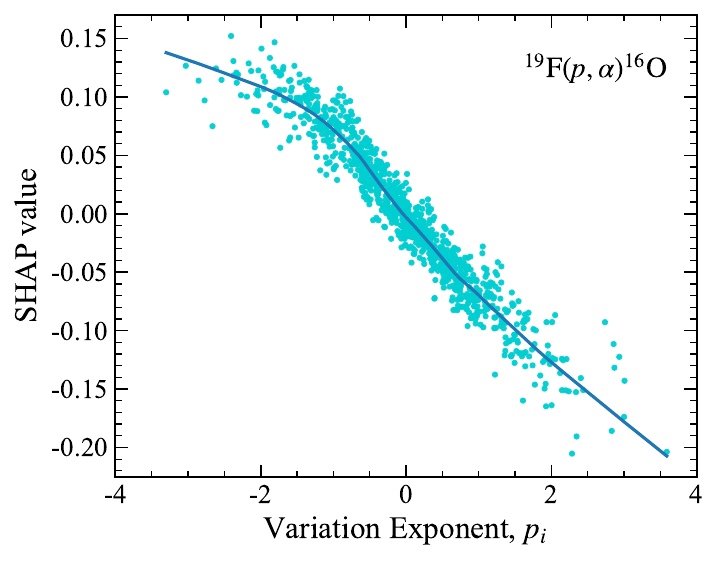}{0.38\textwidth}{}
    \hfill
	}
	\vspace{-31pt}
\caption{\label{fig:F18_and_F19}(Color online) (Left Panels) Effect of $(p, \gamma)$, $(p, \alpha)$ reactions on $^{18}$F, $^{19}$F reaction-rate variations on Ca mass fraction. Overlaying the data points are the 68\% C.I. (red), 95\% C.I.(green) and 99.7\% C.I. (blue) deviations about the mean of the data points. A linear regression is performed on the raw data (gold solid line). The linear regression $R^2$ and Spearman correlation coefficient $r_s$ are also depicted. 
(Right Panels) The dependence of the SHAP values on those reaction-rate variations. The solid blue lines represent the LOWESS (Locally Weighted Scatterplot Smoothing) smoothing fit results.}
\end{figure*}

Figure~\ref{fig:SHAP_values_and_percentages} shows the SHAP summary plots. We quantify the global importance of each reaction rate using the mean absolute SHAP values. We can clearly see that the $(p,\gamma)$ and $(p,\alpha)$ reactions of $^{18}$F play a dominant role in both sets, while the contribution of $^{19}$F$(p,\gamma)^{20}$Ne is reduced in SetB. Figure~\ref{fig:F18_and_F19} in the left panels show the effects of $^{18}$F$(p, \gamma)^{19}$Ne, $^{18}$F$(p, \alpha)^{15}$O, $^{19}$F$(p, \gamma)^{20}$Ne and $^{19}$F$(p, \alpha)^{16}$O reaction-rate variations on Ca mass fraction directly, while the right panels show the dependence of the SHAP values on the reaction-rate variations of these reactions, which clearly capture the contributions of each reaction to the predicted Ca abundance. 

\subsection{\label{subsec:Reaction Interaction}Reaction Interactions}

\begin{figure}[!htbp]
	\centering
    \includegraphics[width=\columnwidth]{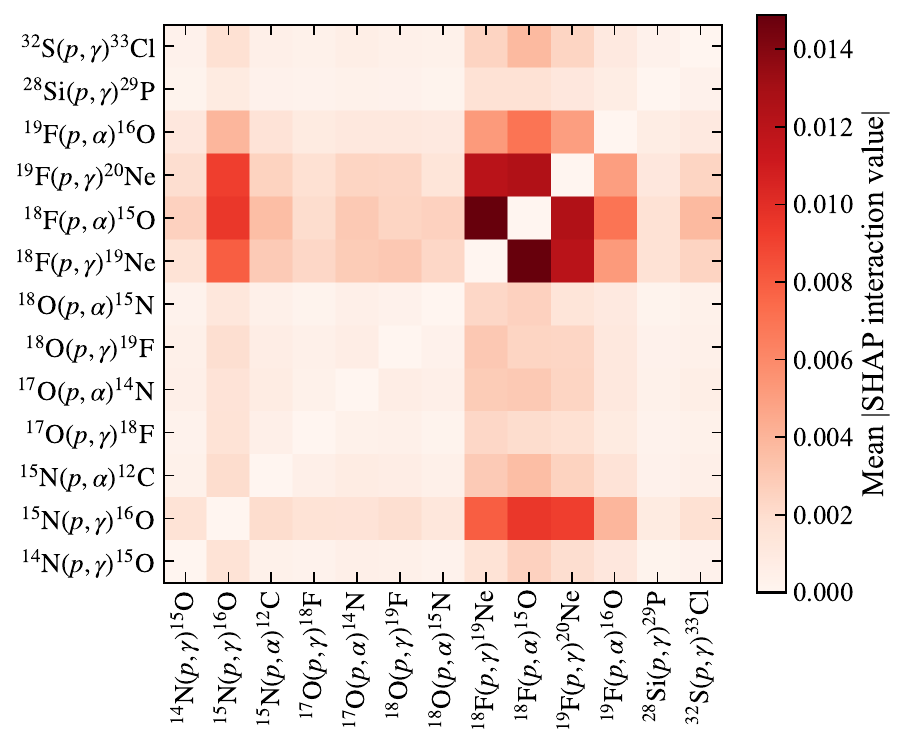}\\
	\includegraphics[width=\columnwidth]{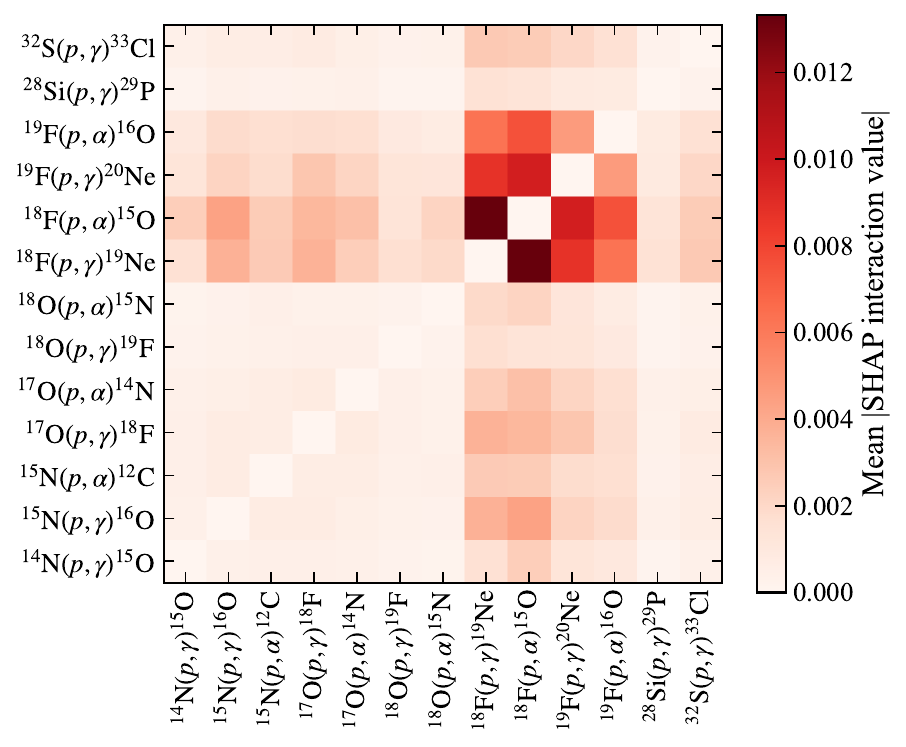}
	\caption{\label{fig:SetB_SHAP_interaction}(Color online) The matrix of mean absolute SHAP interaction values for the 13 reactions for SetA (Top) and SetB (Bottom). The anti-diagonal values of the matrix are set to be zero.}
\end{figure}

SHAP interaction values are used to capture pairwise interaction effects. We can gain additional insight by separating interaction effects from main effects~\citep{Lundberg2019}. Figure~\ref{fig:SetB_SHAP_interaction} illustrates the matrix of mean absolute SHAP interaction values for the 13 reactions for SetA and SetB. Both for SetA and SetB, there are strong interactions among $^{18}$F$(p, \gamma)^{19}$Ne, $^{18}$F$(p, \alpha)^{15}$O, $^{19}$F$(p, \gamma)^{20}$Ne and $^{19}$F$(p, \alpha)^{16}$O reactions. In addition, for SetA, $^{15}$N$(p, \gamma)^{16}$O also have strong interactions with these reactions.

\begin{figure}[!htbp]
	\centering
    \includegraphics[width=\columnwidth]{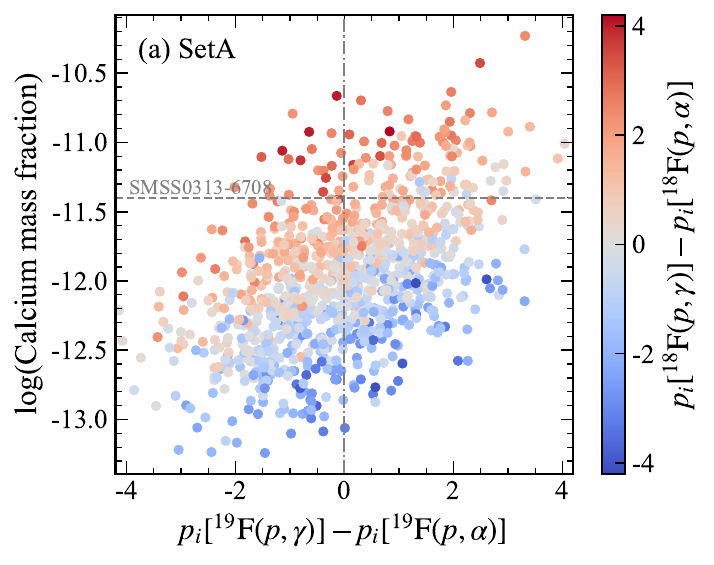}\\
	\includegraphics[width=\columnwidth]{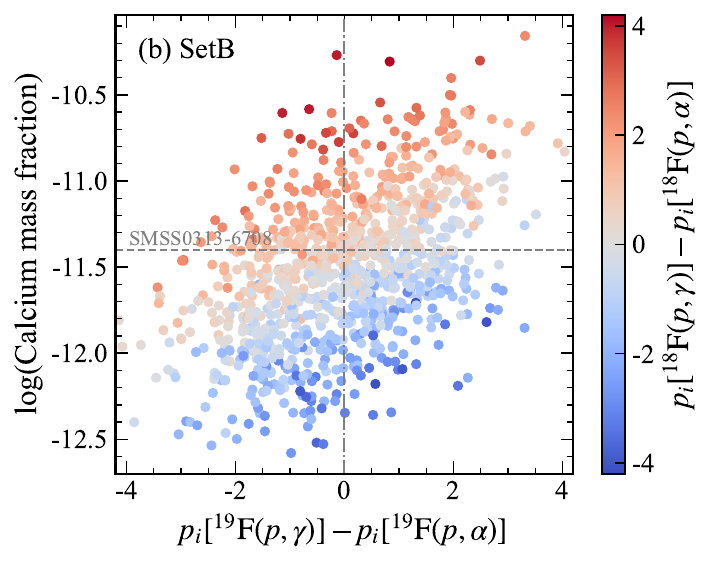}
	\caption{\label{fig:interaction_whole}(Color online) Scatter plots illustrating the sensitivity of the Ca mass fraction to the competition between the $(p,\gamma)$ and $(p,\alpha)$ channels of $^{19}\mathrm{F}$ and $^{18}\mathrm{F}$. The x-axis shows the difference between the variations of the $^{19}\mathrm{F}(p,\gamma)$ and $^{19}\mathrm{F}(p,\alpha)$ reaction rates, $p_i[^{19}\mathrm{F}(p,\gamma)]-p_i[^{19}\mathrm{F}(p,\alpha)]$, while the y-axis represents the Ca mass fraction obtained from  Monte Carlo simulations. The color of the points encodes the corresponding channel competition in $^{18}\mathrm{F}$ between the $(p,\gamma)$ and $(p,\alpha)$, defined as $p_i[^{18}\mathrm{F}(p,\gamma)]-p_i[^{18}\mathrm{F}(p,\alpha)]$. This figure visualizes how the impact of the $^{19}\mathrm{F}$ channel competition on the final yield is modulated by the corresponding reaction-channel competition in $^{18}\mathrm{F}$.}
\end{figure}

In order to investigate the sensitivity of the Ca mass fraction to the competition between the $(p,\gamma)$ and $(p,\alpha)$ channels of $^{19}\mathrm{F}$ and $^{18}\mathrm{F}$, we construct scatter plots using the Monte Carlo samples, as shown in Figure~\ref{fig:interaction_whole}. The x-axis represents the difference between the sampled variations of the $^{19}\mathrm{F}(p,\gamma)$ and $^{19}\mathrm{F}(p,\alpha)$ reaction rates, $p_i[^{19}\mathrm{F}(p,\gamma)]-p_i[^{19}\mathrm{F}(p,\alpha)]$, while the y-axis shows the resulting Ca mass fraction for each sample. The color of each point encodes the corresponding channel competition in $^{18}\mathrm{F}$, defined as $p_i[^{18}\mathrm{F}(p,\gamma)]-p_i[^{18}\mathrm{F}(p,\alpha)]$. This representation clearly illustrates how the effect of $^{19}\mathrm{F}$ reaction-channel competition on the final Ca yield is modulated by the competing reactions in $^{18}\mathrm{F}$. For SetA, if $p_i[^{19}\mathrm{F}(p,\gamma)]-p_i[^{19}\mathrm{F}(p,\alpha)]$=0, the predicted Ca abundance also has the opportunity to reach the observed level in SMSS0313-6708. 

\subsection{\label{subsec:Distribution of Calcium Abundance}Distributions of Calcium Abundances}

\begin{figure*}[!htp]
	\centering
	\gridline{		
		\fig{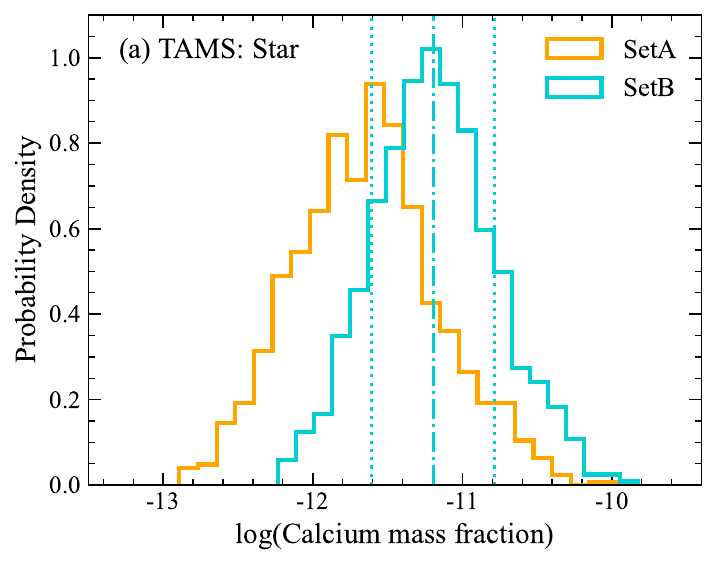}{0.49\textwidth}{}
		\fig{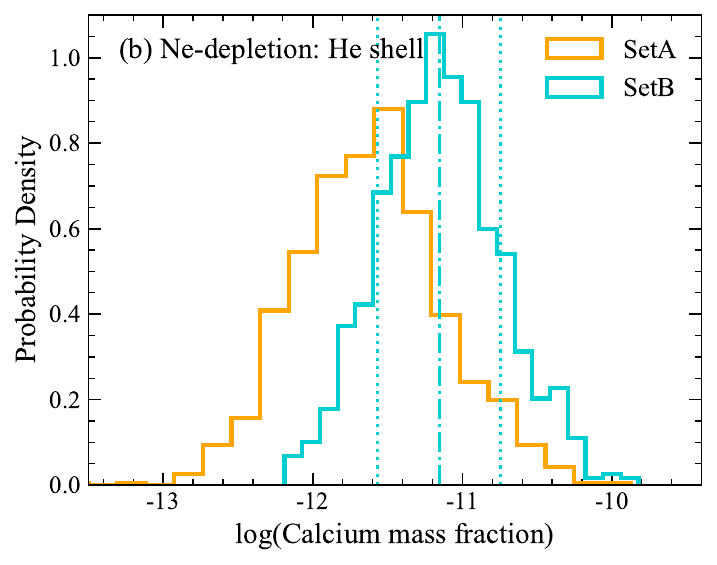}{0.49\textwidth}{}
	}
	\vspace{-30pt}
    \gridline{		
		\fig{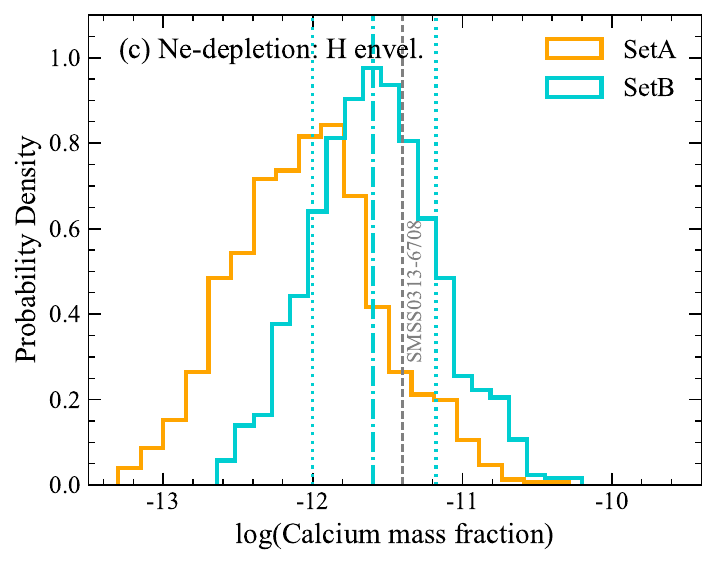}{0.49\textwidth}{}
		\fig{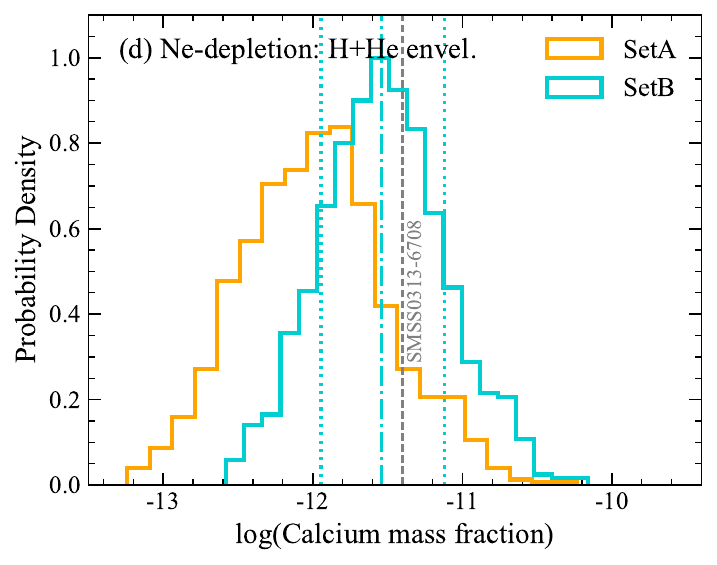}{0.49\textwidth}{}
	}
	\vspace{-30pt}
	\caption{\label{fig:MC_Ca_Distribution}(Color online) Probability density functions for the Ca mass fraction obtained from Monte Carlo simulations for the whole star at the TAMS (a), the He shell ($X(^{4}\mathrm{He}) \ge 0.01$ and $X(^{1}\mathrm{H}) < 0.01$) (b), the H envelope ($X(^{1}\mathrm{H}) \ge 0.01$) (c), and their combination ($X_{\mathrm{b}}(^{4}\mathrm{He})=0.01$ or $X(^{4}\mathrm{He}) \ge 0.01$) (d) at core Ne-depletion. The orange and cyan histograms represent the results of SetA and SetB, respectively. In each panel, the cyan dot-dashed line indicates the median value of the Ca mass fraction, and the cyan dot-dotted lines indicate the 68\% C.I. for SetB.}
\end{figure*}

Figure~\ref{fig:MC_Ca_Distribution} shows the probability density functions (PDFs) for the Ca mass fraction at the TAMS and the core Ne-depletion obtained from Monte Carlo simulations for SetA and SetB. The PDFs comprehensively reflect the uncertainty propagation of nuclear reaction rates and the consistency between model predictions and observational data. They exhibit distinct differences between the two reaction-rate sets. SetA (STARLIB rates) shows a broad distribution centered at a lower values, while SetB shows a narrow distribution centered at a higher values. The distribution of differences between the whole star at the TAMS and the combined H envelope and He shell at Ne depletion is centered at $\sim0.34$ dex for both Set A and Set B, with a narrow 68\% C.I. of $[0.33, 0.35]$ dex. 
  
\begin{figure}[!htbp]
	\centering
	\includegraphics[width=\columnwidth]{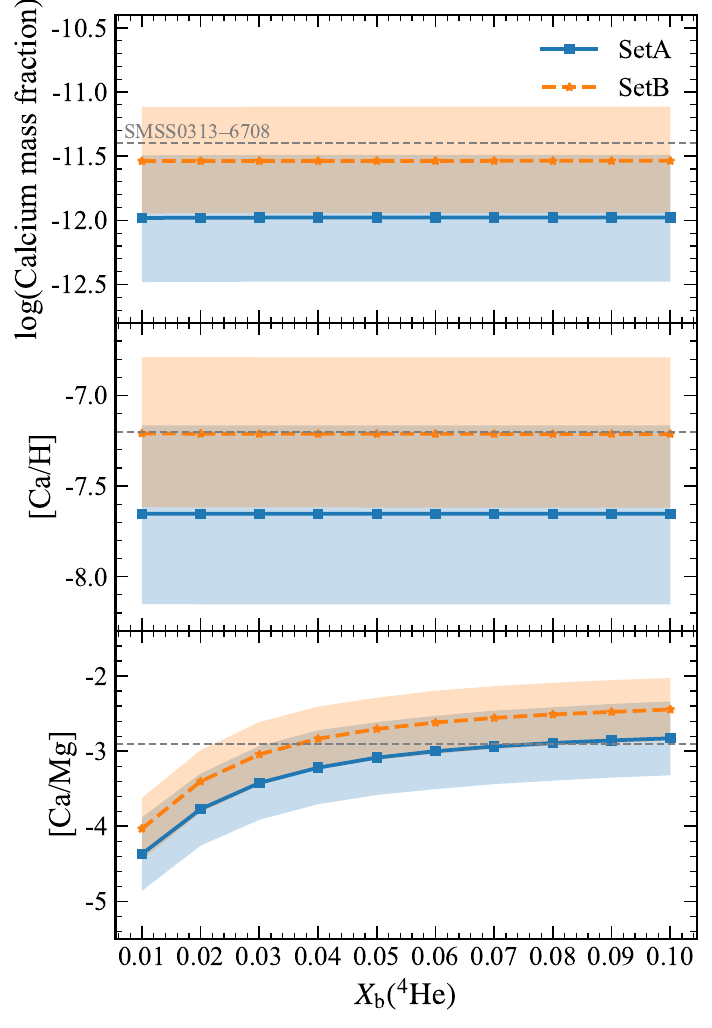}
	\caption{\label{fig:Monte_Carlo_Error_Bar}(Color online) The Ca mass fraction (top), [Ca/H] (middle), and [Ca/Mg] (bottom) as a function of $^4$He mass fraction for MC simulations at core Ne-depletion.The blue solid lines with box markers and orange dashed lines with stars represent the median values of SetA and SetB, respectively. The 68\% C.I. is represented by the shaded bands.}
\end{figure}

In Figure~\ref{fig:Monte_Carlo_Error_Bar}, one can see that the predicted Ca mass fraction, [Ca/H] and [Ca/Mg] agree well with the observed values in SMSS0313–6708 using SetB. In contrast, the observed Ca mass fraction lies marginally outside the predicted 68\% C.I. using SetA. 

\section{\label{sec:Impacts of the updated rates}20\,\Msun and 40\,\Msun Pop III stellar Models}
\begin{figure}[!htbp]
	\centering
	\includegraphics[width=\columnwidth]{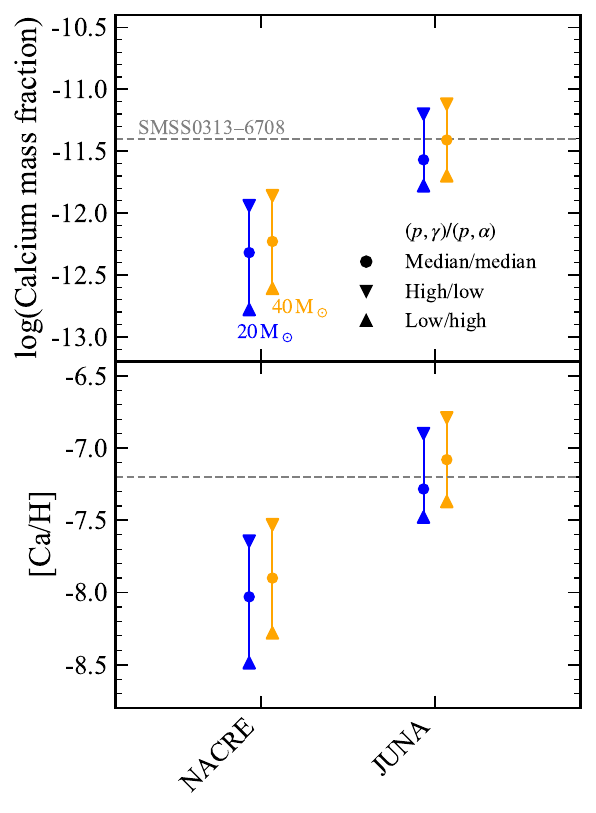}
	\caption{\label{fig:Update_Rate_F19_Ca}(Color online) Prediction of Ca mass fraction and [Ca/H] with different rate sets for 20\,\Msun (blue) and 40\,\Msun (orange) stellar models. See text for details.}
\end{figure}
\begin{figure}[!htbp]
	\centering
	\includegraphics[width=\columnwidth]{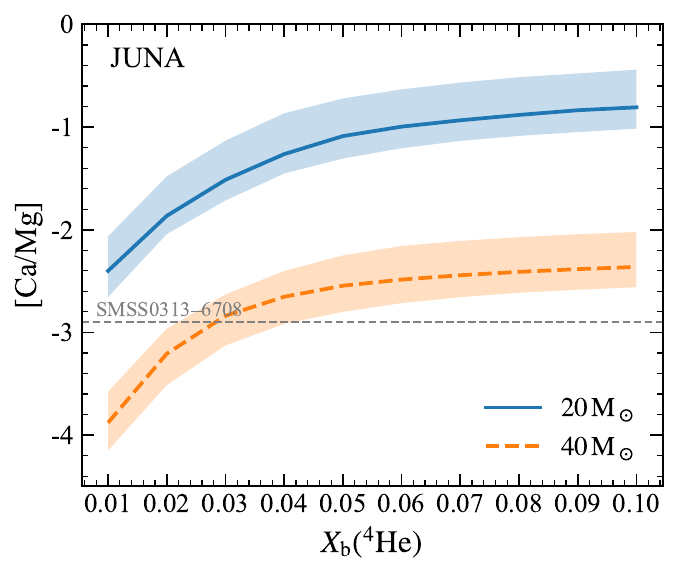}
	\caption{\label{fig:Update_Rate_F19_CaMg}(Color online) The [Ca/Mg] as a function of $^4$He mass fraction at core Ne-depletion for 20\,\Msun (blue) and 40\,\Msun (orange) stellar models using the JUNA $^{19}\mathrm{F}(p, \gamma)^{20}\mathrm{Ne}$ rate. The blue solid and dashed orange lines denote [Ca/Mg] for the combination of the median rates for $^{19}\mathrm{F}(p, \gamma)^{20}\mathrm{Ne}$ and $^{19}\mathrm{F}(p, \alpha)^{16}\mathrm{O}$ rates. See text for details.}
\end{figure}

\citet{Su25} concluded that increasing the $^{19}\mathrm{F}(p,\gamma)/^{19}\mathrm{F}(p,\alpha)$ ratio compared to the NACRE compilation~\citep{Angulo1999} is insufficient to explain the observed Ca abundance ([Ca/Mg] ratio) in stars like SMSS0313–6708, based on their 20\,\Msun and 25\,\Msun Pop III stellar models (see reference therein). 
To respond to this concern, we evolve 20\,\Msun and 40\,\Msun Pop III stellar models using \MESA, following the same procedure of~\citet{Zhang2022}. The Median/median (Mean/mean), High/low, and Low/high combinations of the $^{19}\mathrm{F}(p,\gamma)/^{19}\mathrm{F}(p,\alpha)$ reaction rates allow obtaining the median, upper, and lower limits of Ca mass fraction, [Ca/H], [Ca/Mg] (see the supplementary information of~\citet{Zhang2022} for more details). For the $^{19}$F$(p, \alpha)^{16}$O rate, we use the NACRE~\citep{Angulo1999} rate. For the $^{19}$F$(p, \gamma)^{20}$Ne rate, we use NACRE~\citep{Angulo1999} and JUNA rates~\citep{Zhang2022}. Evolving the 20\,\Msun Pop III stellar models until core Ne-depletion also provides a good approximation. 

In Figure~\ref{fig:Update_Rate_F19_Ca}, we find that Ca mass fraction (see the Table~\ref{tab:ca_production} in Appendix~\ref{sec:Calcium Table}), [Ca/H] obtained both from 20\,\Msun and 40\,\Msun stellar models agree well with corresponding observed values when using the JUNA $^{19}\mathrm{F}(p, \gamma)^{20}\mathrm{Ne}$ rate. However, as shown in Figure~\ref{fig:Update_Rate_F19_CaMg}, only 40\,\Msun stellar models could produce the observed [Ca/Mg] ratio.

\section{\label{sec:Discussion and Conclusion}Discussion and Conclusion}
This study advances our understanding of calcium nucleosynthesis in Population III stars through three key contributions: (1) a systematic identification of 13 key reactions regulating calcium production through 1D multi-zone models, (2) quantification of reaction interactions and uncertainty propagation to the abundance of calcium via Monte Carlo simulations, and (3) integration of Spearman rank-order correlation and SHAP values to interpret the complex relationships between reaction rates and calcium synthesis.

Sensitivity tests identified 13 reactions that significantly impact calcium abundance, with the \((p, \gamma)\) and \((p, \alpha)\) reactions of \(^{18}\text{F}\) and \(^{19}\text{F}\) emerging as dominant contributors. Spearman correlation analysis revealed strong monotonic relationships between these reactions and calcium production (\(|r_s| > 0.25\), confirming their critical role in regulating reaction flows toward calcium formation. Notably, \(^{18}\text{F}(p, \alpha)^{15}\text{O}\) exhibited the strongest negative correlation (\(r_s = -0.62\)), while \(^{18}\text{F}(p, \gamma)^{19}\text{Ne}\) showed a robust positive correlation (\(r_s = 0.51\)), highlighting their competing effects on channel leakage from the CNO cycle to calcium-producing pathways.

SHAP analysis further quantified the relative importance of these reactions, with \(^{18}\text{F}(p, \alpha)^{15}\text{O}\) and \(^{18}\text{F}(p, \gamma)^{19}\text{Ne}\) accounting for \(\sim28\%\) and \(\sim24\%\) of the total model output variation, respectively. This aligns with previous findings that \(^{19}\text{F}(p, \gamma)^{20}\text{Ne}\) acts as a key breakout reaction from the CNO cycle, facilitating irreversible flow toward heavier isotopes like \(^{40}\text{Ca}\) \citep{Zhang2022}.

Monte Carlo simulations propagated these uncertainties, showing that SetB (updated rates) achieves better agreement with SMSS0313–6708's calcium abundance within the 68\% C.I., whereas SetA (STARLIB rates) marginally falls outside this range.

Comparisons between 20\,\Msun and 40\,\Msun models showed that the 40\,\Msun framework produces calcium abundances more consistent with observations, particularly when using updated JUNA rates for \(^{19}\text{F}(p, \gamma)^{20}\text{Ne}\). The 40\,\Msun model's extended hydrogen-burning phase and higher central temperatures likely enhance CNO cycle breakout efficiency, supporting previous proposals that intermediate-mass Population III stars are key contributors to calcium enrichment in the early universe \citep{Takahashi2014}. Core Ne-depletion emerged as a reliable reference point for comparing model predictions with observations, as calcium abundances remained stable between this stage and pre-supernova collapse.

Our findings consistent with \citet{Zhang2022}, we confirm that \(^{19}\text{F}(p, \gamma)^{20}\text{Ne}\) is a critical breakout reaction from the CNO cycle, with its rate enhancement (by a factor of 5.4–7.4 at 0.1 GK) improving agreement with SMSS0313–6708's calcium abundance. However, our results contradict \citet{Su25}, who argued that increasing the \(^{19}\text{F}(p, \gamma)/^{19}\text{F}(p, \alpha)\) ratio is insufficient to explain observed calcium levels.

Additionally, our Monte Carlo approach advances beyond the post-processing schemes of \citet{Fields2016, Fields2018} by directly propagating reaction-rate uncertainties within stellar evolution models, enabling more realistic comparisons with observational constraints.

\begin{acknowledgments}
The authors thank the reviewer for her/his valuable comments on the submitted manuscript. Special thanks to Jun Hu for providing the data of $^{18}$O$(p, \alpha)^{15}$N  in his published APJ paper. We acknowledge support from the National Natural Science Foundation of China (No. 12175152), LingChuang Research Project of China National Nuclear Corporation (Nos. 2025-084, 2024-065, 2024-068).
\end{acknowledgments}

\software{\MESA~\citep[r24.03.01,][]{Paxton2010,Paxton2013,Paxton2015,Paxton2018,Paxton2019,Jermyn2023}, \texttt{IPython}~\citep{ipython}, \texttt{Jupyter}~\citep{jupyter}, \texttt{XGBoost}~\citep{Chen2016}, \texttt{SHAP}~\citep{Lundberg2017,Lundberg2020}, \texttt{matplotlib}~\citep{matplotlib}, \texttt{numpy}~\citep{numpy}, \texttt{scipy}~\citep{2020SciPy-NMeth}, \texttt{scikit-learn}~\citep{scikit-learn}, \texttt{seaborn}~\citep{Waskom2021}}

\clearpage                    
\appendix
\section{\label{sec:Calcium Table}Calcium Abundance} 
\startlongtable
\begin{deluxetable*}{cccccc}
\tablecaption{\label{tab:ca_sensitivity}
Average calcium mass fraction for different reaction-rate variations by a factor of 10 up and down in 40\,\Msun Pop III stellar models}
\tablehead{
\colhead{} &
\colhead{} &
\multicolumn{1}{c}{TAMS} &
\multicolumn{3}{c}{Ne-depletion} \\
\cline{3-3} \cline{4-6}
\colhead{Reaction}&\colhead{Variation} &
\colhead{Star} &
\colhead{H envel.} &
\colhead{He shell} &
\colhead{H+He envel.}
}
\startdata
Baseline      & ... & $-11.86$ & $-12.26$ & $-11.82$ & $-12.20$ \\
$\isotope[14][]{N}(p,\gamma)\isotope[15][]{O}$ & Down & $-11.92$ & $-12.32$ & $-11.83$ & $-12.25$ \\
               & Up   & $-11.84$ & $-12.24$ & $-12.05$ & $-12.22$ \\
$\isotope[15][]{N}(p,\gamma)\isotope[16][]{O}$ & Down & $-12.86$ & $-13.26$ & $-12.82$ & $-13.20$ \\
              & Up   & $-10.87$ & $-11.26$ & $-10.84$ & $-11.21$ \\
$\isotope[15][]{N}(p,\alpha)\isotope[12][]{C}$ & Down & $-10.87$ & $-11.26$ & $-10.82$ & $-11.21$ \\
              & Up   & $-12.86$ & $-13.26$ & $-12.83$ & $-13.20$ \\
$\isotope[17][]{O}(p,\gamma)\isotope[18][]{F}$ & Down & $-12.86$&$-13.26$&$-12.81$&$-13.20$  \\
               & Up   & $-10.89$&$-11.29$& $-10.86$& $-11.23$  \\
$\isotope[17][]{O}(p,\alpha)\isotope[14][]{N}$ & Down & $-10.88$&$-11.28$&$-10.85$&$-11.23$  \\
              & Up   & $-12.86$ &$-13.26$&$-12.83$&$-13.20$  \\
$\isotope[18][]{O}(p,\gamma)\isotope[19][]{F}$ & Down & $-12.14$ &$-12.51$&$-12.11$&$-12.46$  \\
              & Up   & $-11.13$ &$-11.55$&$-11.10$&$-11.49$  \\
$\isotope[18][]{O}(p,\alpha)\isotope[15][]{N}$ & Down & $-11.13$&$-11.55$&$-11.09$&$-11.49$  \\
               & Up   & $-12.14$ &$-12.51$&$-12.10$&$-12.46$ \\
$\isotope[18][]{F}(p,\gamma)\isotope[19][]{Ne}$ & Down & $-12.11$&$-12.53$ &$-12.07$&$-12.47$  \\
              & Up   & $-11.14$&$-11.51$ &$-11.11$&$-11.46$  \\
$\isotope[18][]{F}(p,\alpha)\isotope[15][]{O}$ & Down & $-11.10$ &$-11.47$&$-11.07$&$-11.42$  \\
              & Up   & $-12.76$&$-13.17$&$-12.72$&$-13.11$  \\
$\isotope[19][]{F}(p,\gamma)\isotope[20][]{Ne}$ & Down & $-12.86$ & $-13.26$ &$-12.84$ & $-13.21$  \\
               & Up   & $-10.87$ &$-11.26$&$-10.83$&$-11.21$ \\
$\isotope[19][]{F}(p,\alpha)\isotope[16][]{O}$ & Down & $-10.87$ &$-11.27$ &$-10.84$&$-11.21$  \\
              & Up   & $-12.86$ &$-13.26$ &$-12.83$&$-13.20$  \\
$\isotope[28][]{Si}(p,\gamma)\isotope[29][]{P}$ & Down & $-11.96$ &$-12.35$ &$-11.93$&$-12.30$  \\
              & Up   & $-11.85$ &$-12.25$ &$-11.82$&$-12.19$  \\
$\isotope[32][]{S}(p,\gamma)\isotope[33][]{Cl}$ & Down & $-12.41$ &$-12.80$ &$-12.38$ &$-12.75$  \\
              & Up   & $-11.70$ &$-12.13$ &$-11.65$ &$-12.06$  \\          
\enddata
\tablecomments{He shell ($X(^{4}\mathrm{He}) \ge 0.01$ and $X(^{1}\mathrm{H}) < 0.01$), H envelope ($X(^{1}\mathrm{H}) \ge 0.01$), and their combination ($X_{\mathrm{b}}(^{4}\mathrm{He})=0.01$ or $X(^{4}\mathrm{He}) \ge 0.01$). Values are given in logarithm base 10.}
\end{deluxetable*}

\startlongtable
\begin{deluxetable*}{cccccccc}
\tablecaption{\label{tab:ca_production}
Comparison of average calcium mass fraction for various reaction-rate sets in 20\,\Msun and 40\,\Msun Pop III stellar models}
\tablehead{
\multicolumn{2}{c}{}&
\multicolumn{2}{c}{}&
\multicolumn{1}{c}{TAMS, 20\,\Msun(40\,\Msun)} &
\multicolumn{3}{c}{Ne-depletion, 20\,\Msun(40\,\Msun)} \\
\cline{5-5} \cline{6-8}
\multicolumn{2}{c}{$\isotope[19][]{F}(p,\gamma)\isotope[20][]{Ne}$}&
\multicolumn{2}{c}{$\isotope[19][]{F}(p,\alpha)\isotope[16][]{O}$}& 
\colhead{Star} &
\colhead{H envel.} &
\colhead{He shell} &
\colhead{H+He envel.}
}
\startdata
     &Low&     &High&$-12.33(-12.26)$&$-12.82(-12.66)$&$-12.51(-12.23)$&$-12.78(-12.61)$\\
NACRE&Median&NACRE&Median&$-11.95(-11.89)$&$-12.46(-12.29)$&$-11.76(-11.85)$&$-12.32(-12.23)$\\
     &High&    &Low&$-11.57(-10.89)$&$-12.09(-11.91)$&$-11.37(-11.48)$&$-11.94(-11.86)$\\   
     &Low&     &High&$-11.42(-11.36)$&$-11.93(-11.76)$&$-11.19(-11.32)$&$-11.78(-11.70)$\\
JUNA&Median&NACRE&Median&$-11.10(-11.07)$&$-11.60(-11.46)$&$-11.32(-11.03)$&$-11.57(-11.41)$\\   
     &High&    &Low&$-10.79(-10.77)$&$-11.30(-11.17)$&$-11.70(-10.75)$&$-11.20(-11.12)$\\     
\enddata
\tablecomments{He shell ($X(^{4}\mathrm{He}) \ge 0.01$ and $X(^{1}\mathrm{H}) < 0.01$), H envelope ($X(^{1}\mathrm{H}) \ge 0.01$), and their combination ($X_{\mathrm{b}}(^{4}\mathrm{He})=0.01$ or $X(^{4}\mathrm{He}) \ge 0.01$). Values are given in logarithm base 10.}
\end{deluxetable*}

\section{\label{sec:Variations of Reaction rates}Spearman correlation coefficient and SHAP value} 
\begin{figure*}[!htbp]
\centering
	\gridline{		
		\fig{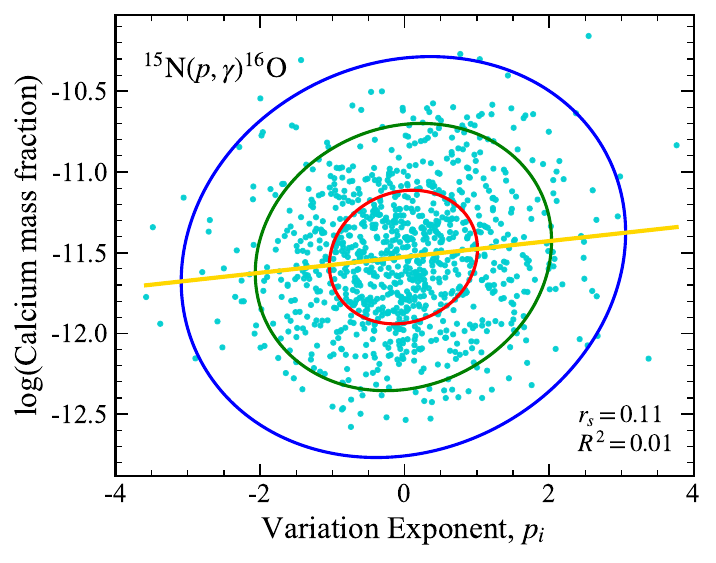}{0.40\textwidth}{}
		\fig{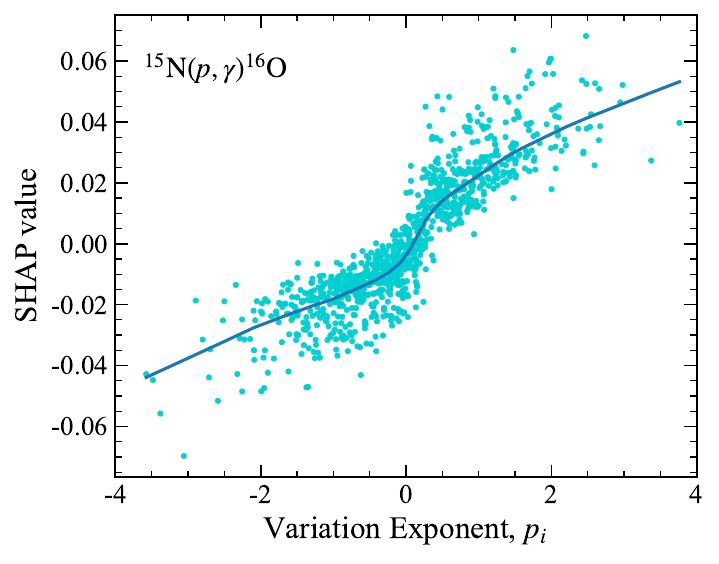}{0.40\textwidth}{}
	}
	\vspace{-25pt}
    \gridline{		
		\fig{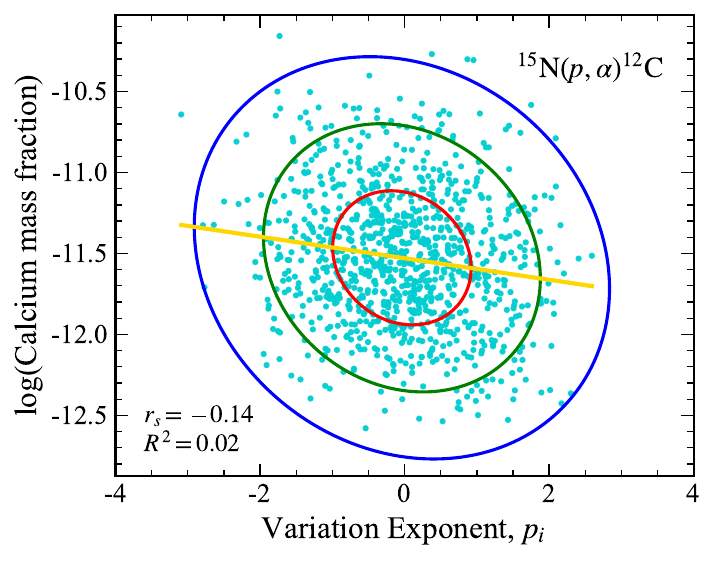}{0.40\textwidth}{}
		\fig{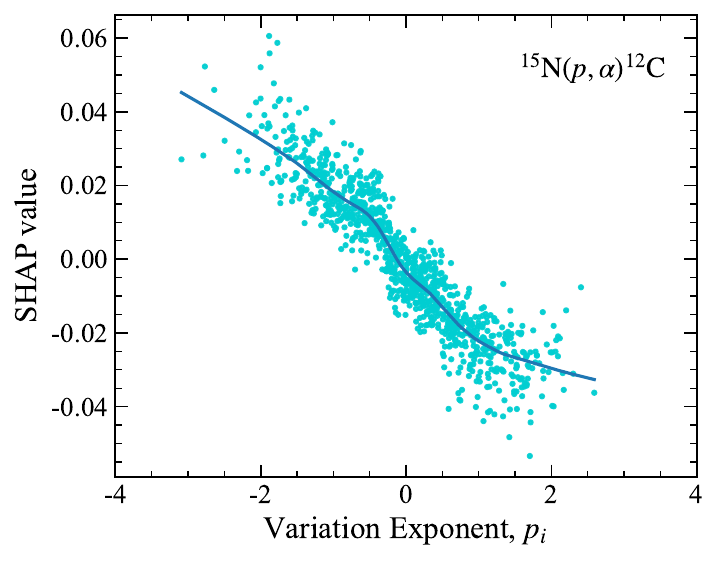}{0.40\textwidth}{}
	}
\caption{\label{fig:N15}(Color online) Same as Figure \ref{fig:F18_and_F19} but for $^{15}$N$(p, \gamma)^{16}$O and $^{15}$N$(p, \alpha)^{12}$C reactions (SetB).}
\end{figure*}

\begin{figure*}[!htbp]
\centering
	\gridline{		
		\fig{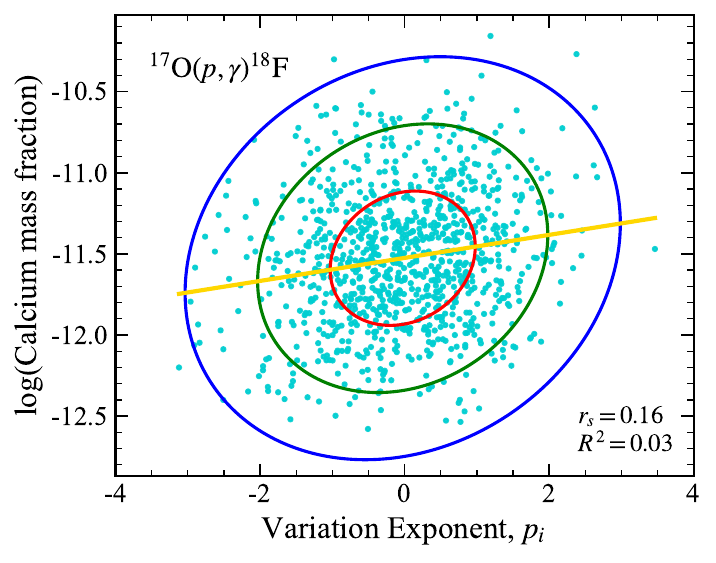}{0.4\textwidth}{}
		\fig{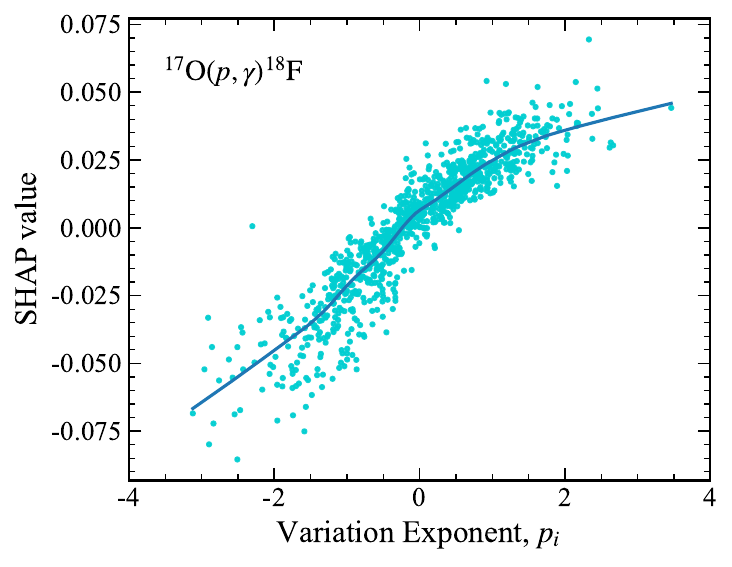}{0.4\textwidth}{}
	}
	\vspace{-30pt}
    \gridline{		
		\fig{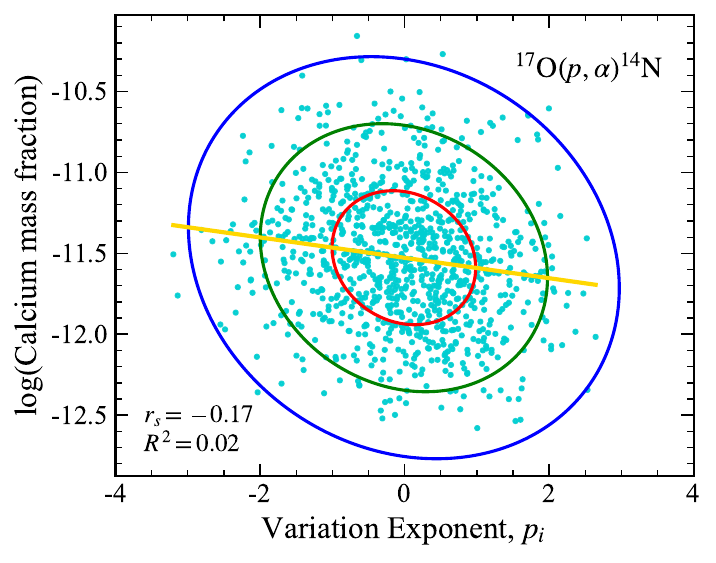}{0.4\textwidth}{}
		\fig{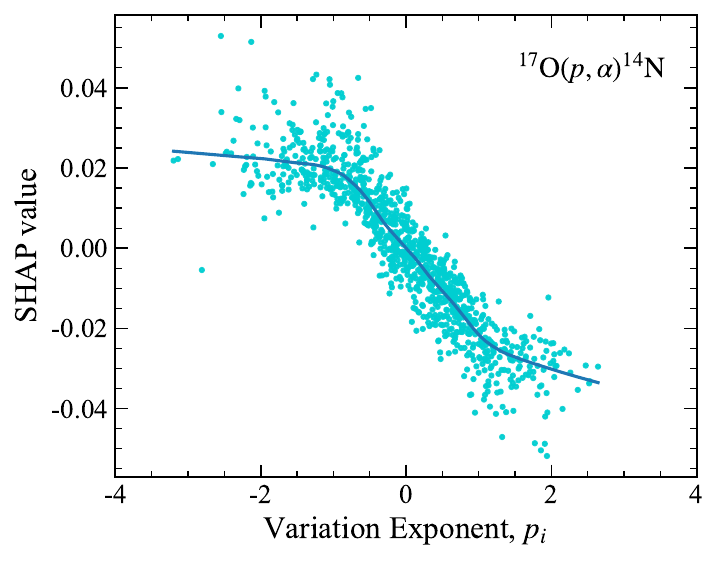}{0.4\textwidth}{}
	}
	\vspace{-30pt}
	\gridline{		
		\fig{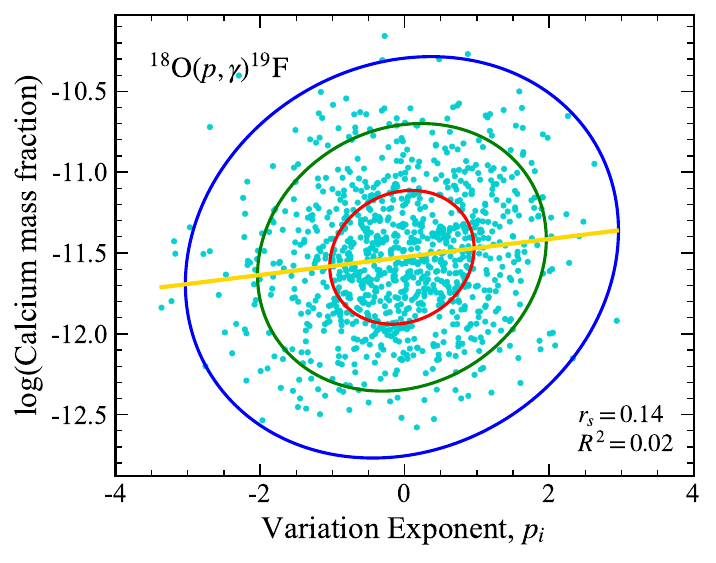}{0.4\textwidth}{}
		\fig{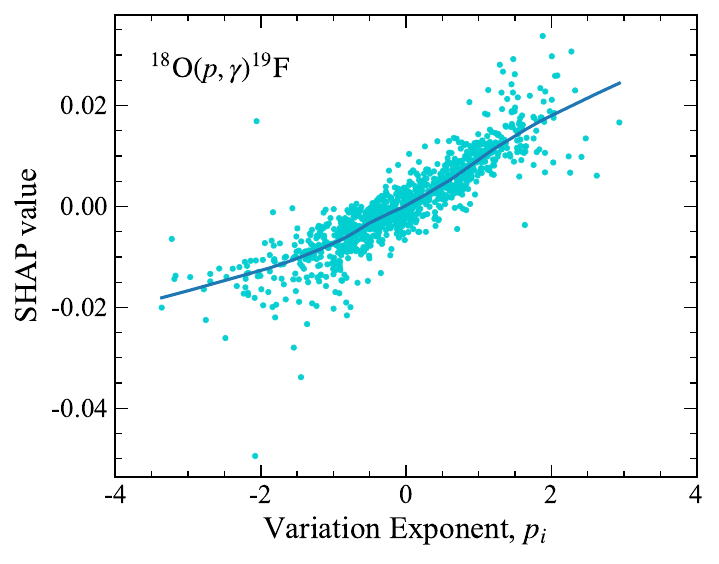}{0.4\textwidth}{}
	}
	\vspace{-30pt}
    \gridline{		
		\fig{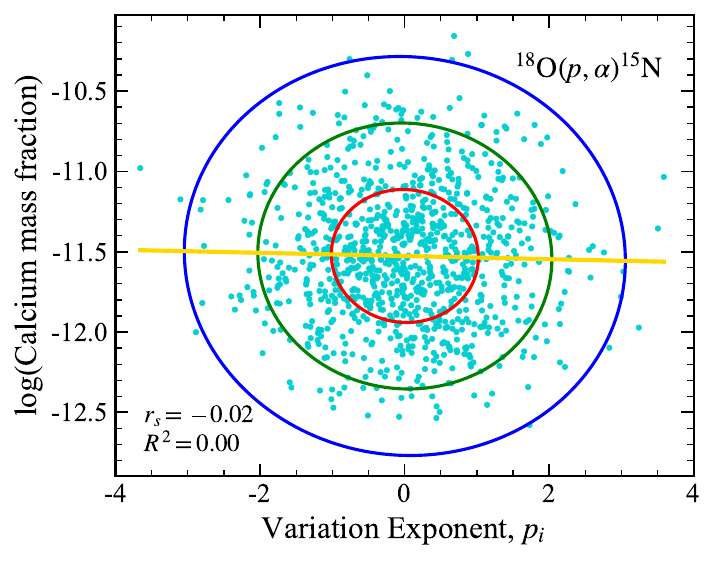}{0.4\textwidth}{}
		\fig{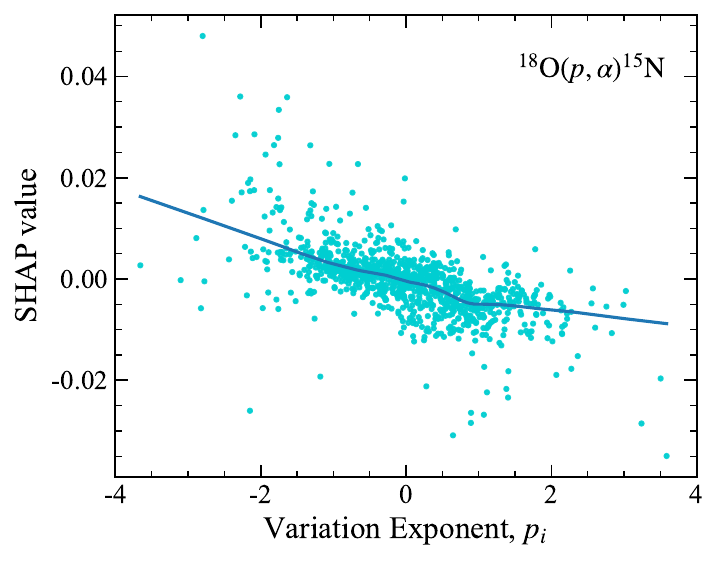}{0.4\textwidth}{}
	}
	\vspace{-30pt}
\caption{\label{fig:O17}(Color online) Same as Figure \ref{fig:F18_and_F19} but for $^{17}$O$(p, \gamma)^{18}$F, $^{17}$O$(p, \alpha)^{14}$N $^{18}$O$(p, \gamma)^{19}$F and	$^{18}$O$(p, \alpha)^{15}$N reactions (SetB).}
\end{figure*}

\begin{figure*}[!htbp]
\centering
    \gridline{		
		\fig{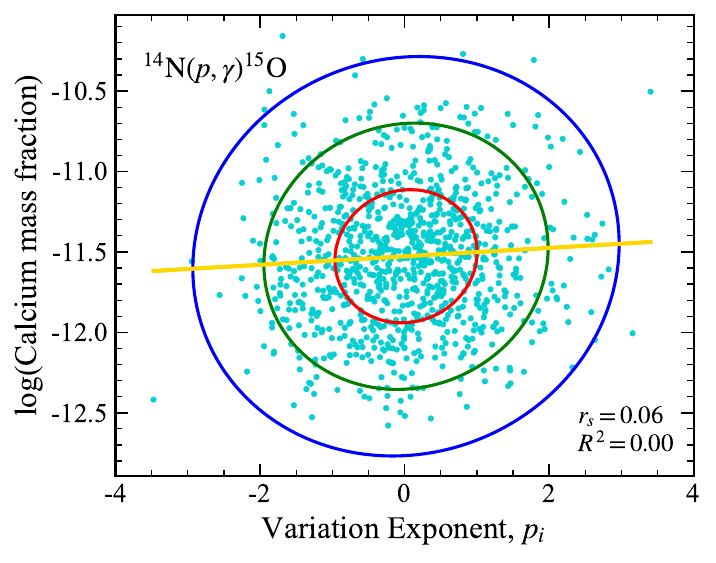}{0.4\textwidth}{}
		\fig{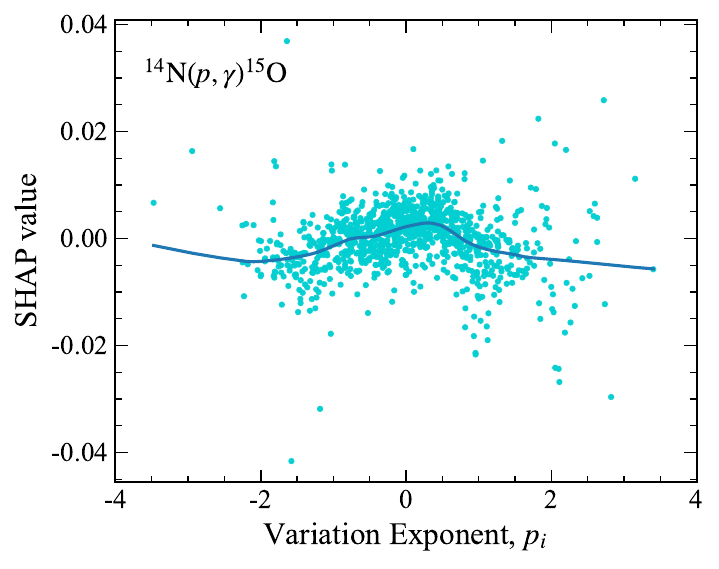}{0.40\textwidth}{}
	}
	\vspace{-25pt}
	\gridline{		
		\fig{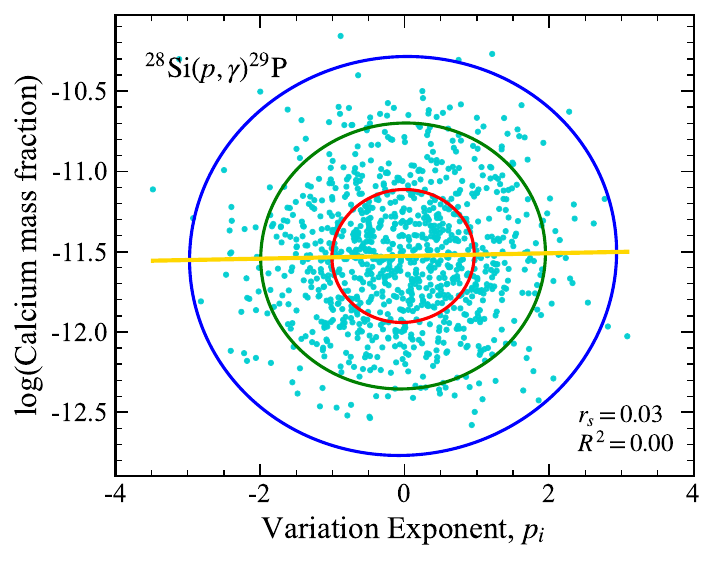}{0.4\textwidth}{}
		\fig{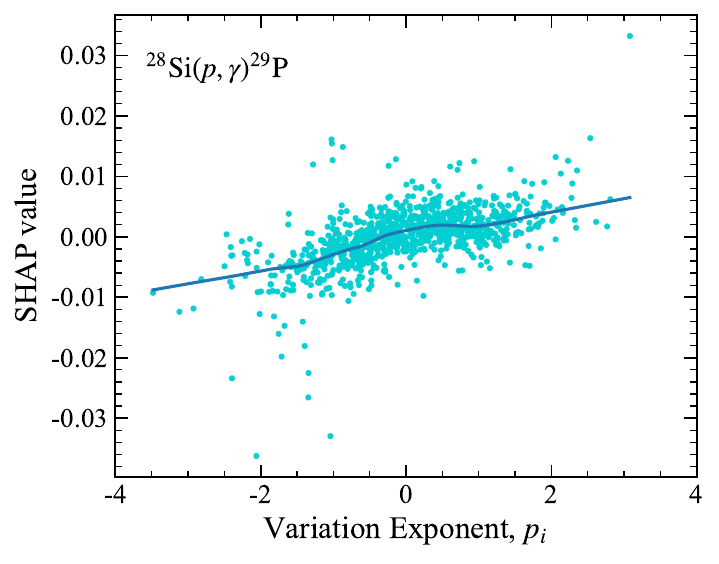}{0.4\textwidth}{}
	}
	\vspace{-25pt}
        \gridline{		
		\fig{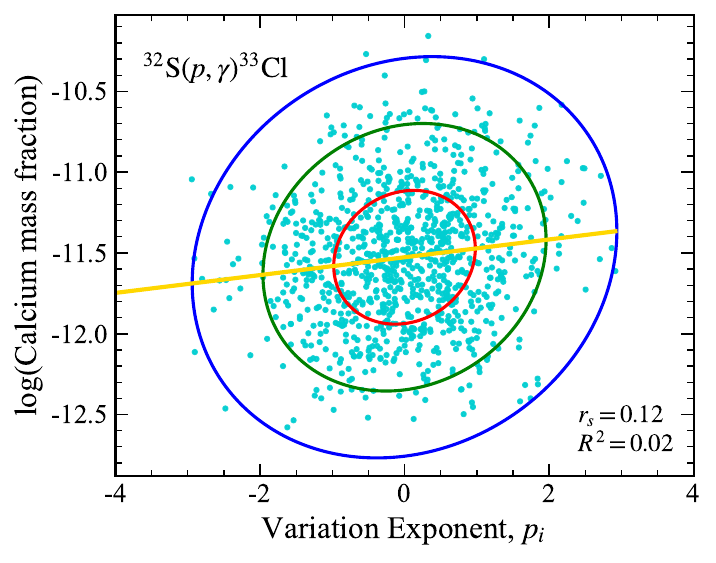}{0.4\textwidth}{}
		\fig{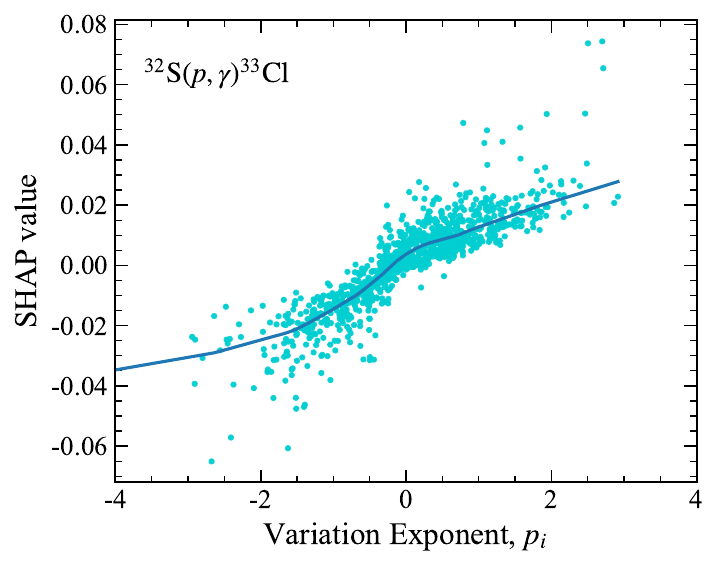}{0.4\textwidth}{}
	}
	\vspace{-25pt}
\caption{\label{fig:NSiS}(Color online) Same as Figure \ref{fig:F18_and_F19} but for $^{14}$N$(p, \gamma)^{15}$O, $^{28}$Si$(p, \gamma)^{29}$P and $^{32}$S$(p, \gamma)^{33}$Cl reactions (SetB).}
\end{figure*}
\clearpage
\bibliography{reference}
\bibliographystyle{aasjournalv7}
\end{document}